\documentclass[showpacs,aps,floatfix,prb,reprint,superscriptaddress]{revtex4-1} 
\usepackage{xfrac}
\usepackage{amssymb}
\usepackage{braket}
\usepackage{graphicx}
\usepackage{verbatim}
\usepackage{etoolbox}
\usepackage{amsfonts} %maths
\usepackage{amsmath} %maths
\usepackage[utf8]{inputenc} %useful to type directly diacritic characters
\usepackage{mathtools}
\usepackage{soul,xcolor,colortbl}
\usepackage{hyperref} \hypersetup{colorlinks=true,citecolor=blue,linkcolor=blue,urlcolor=blue} % HYPERLINKS
\usepackage{bm} %revtex4.1 for bold symbols
\usepackage{array}
\newcolumntype{C}[1]{>{\centering\arraybackslash}p{#1}}\usepackage{soul}
\definecolor{Gray}{gray}{0.85}

\usepackage{color} % STEFANO
\definecolor{Gray}{gray}{0.9}
\definecolor{LightCyan}{rgb}{0.88,1,1}
\def\AGL{{\footnotesize \mathrm{AGL}}}
\def\EXP{{\mathrm{exp}}}
\def\Debye{{\mathrm{D}}}
\def\acoustic{{\mathrm{a}}}
\def\anh{{\mathrm{anh}}}
\def\ML{{\mathrm{ML}}}
\def\WmK{{\footnotesize (W/(m$\cdot$K))}}
\def\K{{\footnotesize (K)}}
\def\AGL{{\small AGL}}

\begin{document}
% \title{\Large AFLOW GIBBS}
\title{\Large High-Throughput Computational Screening of thermal conductivity, Debye temperature and Gr{\"u}neisen parameter \\ using a quasi-harmonic Debye Model}

\author{Cormac Toher}
\affiliation{Department of Mechanical Engineering and Materials Science, Duke University, Durham, North Carolina 27708, USA}
\author{Jose J. Plata}
\affiliation{Department of Mechanical Engineering and Materials Science, Duke University, Durham, North Carolina 27708, USA}
\author{Ohad Levy}
\altaffiliation{On leave from the Department of Physics, NRCN, Israel}
\affiliation{Department of Mechanical Engineering and Materials Science, Duke University, Durham, North Carolina 27708, USA}
\author{Maarten de Jong}
\affiliation{Department of Materials Science and Engineering, University of California, Berkeley, 210 Hearst Memorial Mining Building, Berkeley, USA}
\author{Mark Asta}
\affiliation{Department of Materials Science and Engineering, University of California, Berkeley, 210 Hearst Memorial Mining Building, Berkeley, USA}
\author{Marco Buongiorno Nardelli}
\affiliation{Department of Physics and Department of Chemistry, University of North Texas, Denton TX }
\author{Stefano Curtarolo}
\email[]{stefano@duke.edu}
\affiliation{Materials Science, Electrical Engineering, Physics and Chemistry, Duke University, Durham NC, 27708}

\date{\today}

\begin{abstract}
 The quasi-harmonic Debye approximation has been implemented within the {\small AFLOW} and Materials Project frameworks for high-throughput
 computational materials science (\underline{A}utomatic \underline{G}ibbs \underline{L}ibrary, {\small AGL}), in order to calculate thermal
 properties such as the Debye temperature and the thermal conductivity of materials.
 We demonstrate that the \AGL\ method, which is significantly cheaper computationally compared to the fully {\it ab initio}
 approach, can reliably predict the ordinal ranking of the thermal
 conductivity for several different classes of semiconductor
 materials. In particular, a high Pearson (i.e. linear)
 correlation is obtained between the experimental and 
 {\small AGL} computed values of the lattice thermal conductivity for
 a set of 75 compounds
including materials with cubic, hexagonal,
 rhombohedral and tetragonal symmetry. 
\end{abstract}

\maketitle

\section{Introduction}

Calculating the thermal properties of materials is important for predicting the thermodynamic stability of structural
phases and assessing their importance for a variety of applications.
The lattice thermal conductivity, $\kappa_{l}$, is a crucial design
parameter in a wide range of important technologies,
such us the development of new thermoelectric materials \cite{zebarjadi_perspectives_2012,curtarolo:art84},
heat sink materials for thermal management in electronic devices \cite{Yeh_2002},
and rewritable density scanning-probe phase-change memories \cite{Wright_tnano_2011}.
High thermal conductivity materials, which typically have a zincblende
or diamond-like structure, are essential
in microelectronic and nanoelectronic devices 
to achieve efficient heat removal \cite{Watari_MRS_2001}, and have
been intensively studied for the past few
decades \cite{Slack_1987}.
Low thermal conductivity materials constitute the basis of a new
generation of thermoelectric materials and thermal barrier coatings
\cite{Snyder_jmatchem_2011}. 

The determination of the thermal conductivity of
materials is computationally demanding
since it requires calculation of multiple-phonon scattering processes, that are
the origin of the lattice resistance to heat transport.
The methods most commonly used currently to calculate the thermal conductivity are based on solving the Boltzmann Transport Equation
(BTE). This solution involves the calculation of the phonon frequencies, group velocities, and the harmonic and anharmonic interatomic force constants (IFCs) \cite{Broido2007, Wu_PRB_2012}.
In particular, the third-order anharmonic IFCs are required in order to incorporate the effects of three phonon scattering processes \cite{Broido2007, Wu_PRB_2012}.
The standard method to calculate these anharmonic IFCs is based on density functional theory (DFT). Deinzer et al. \cite{Deinzer_PRB_2003} used Density Functional
Perturbation Theory (DFPT) to obtain third-order IFCs to study the phonon linewidths.
In the last decade, this method has been successfully used to solve the BTE and predict
the thermal conductivity of different materials \cite{Broido2007,ward_ab_2009,ward_intrinsic_2010,Wu_PRB_2012,Zhang_JACS_2012,Li_PRB_2012,Lindsay_PRL_2013,Lindsay_PRB_2013}.
Such evaluation of the higher-order IFCs requires electronic structure calculations for multiple large supercells, each of which has a different set of atomic displacements.
These first principles solutions of the BTE are therefore  computationally extremely expensive.

A variety of simple methods have been devised to evaluate the thermal properties of materials at reduced computational cost.
Early approximate implementations to compute the lattice thermal conductivity were based on semi-empirical models to solve the BTE, in which some parameters are obtained from fitting to experimental
data \cite{ziman_electrons_2001,callaway_model_1959,Allen_PHMB_1994}.
This reduces the predictive power of the calculations. 

An alternative approach to calculating thermal conductivity is based on the Green-Kubo formula, which employs molecular dynamics simulations to calculate thermal currents over long
time periods after thermal equilibrium is reached \cite{Green_JCP_1954,Kubo_JPSJ_1957}. 
This technique takes into account high order scattering processes, but the usage of semi-empirical potentials
leads to errors on the order of 50$\%$ \cite{zebarjadi_perspectives_2012}. 

The methods described above are  unsuitable for rapid generation and screening of large databases of materials properties in order to identify trends and simple
descriptors for thermal properties \cite{curtarolo:art81}.
To accomplish this, we chose to implement a much cheaper approach, the ``GIBBS'' quasi-harmonic Debye model \cite{Blanco_CPC_GIBBS_2004}. This approach does not require large supercell calculations
since it relies merely on first-principles calculations of the energy as a function of unit cell volume. It is thus much more tractable computationally and is eminently suited to
investigating the thermal properties of entire classes of materials in a highly-automated fashion, in order to identify promising candidates for more in-depth
experimental and computational analysis.
We incorporate this model in a new software library, the Automatic GIBBS Library (\AGL), within the {\small AFLOW} \cite{curtarolo:art65,aflowlibPAPER,curtarolo:art92} and
Materials Project \cite{materialsproject.org,APL_Mater_Jain2013,CMS_Ong2012b} frameworks for high-throughput computational materials
science, and use it to construct a database of computed compound thermal conductivities and Debye temperatures.

\section{The Automatic GIBBS Library (AGL)}

The \AGL\ software library implements the ``GIBBS'' method \cite{Blanco_CPC_GIBBS_2004} in the {\small AFLOW} \cite{curtarolo:art65,aflowlibPAPER,curtarolo:art92} framework (C++ based framework) and the
Materials Project \cite{materialsproject.org,APL_Mater_Jain2013,CMS_Ong2012b} (Python implementation). 
The library includes automatic error handling and correction to facilitate high-throughput
computation of materials thermal properties. The principal ingredients of the calculation are described in the following sections. 

\subsection{The GIBBS quasi-harmonic Debye model}

In thermodynamics, the equilibrium state of a system at a constant temperature and pressure minimizes its Gibbs free energy
\begin{equation}
 G(\vec{x}; p, T) = E(\vec{x}) + p V(\vec{x}) + A_{vib}(\vec{x}; T),
\end{equation}
where $\vec{x}$ is a configuration vector containing all the information about the system's geometry, $E(\vec{x})$ is the total energy of the crystal 
(obtained, for example, from an electronic structure calculation), $A_{vib}$ is the vibrational Helmholtz free energy, and $p$ and $V(\vec{x})$ are the pressure and volume. It is assumed here that
the electronic and intrinsic point defect contributions to the Helmholtz free energy is small, which is a good approximation for most materials at temperatures
significantly below their melting point. In the quasi-harmonic approximation, the Helmholtz vibrational energy is
\begin{equation}
 A_{vib}(\vec{x}; T) \!=\!\! \int_0^{\infty} \!\!\left[\frac{\hbar \omega}{2} \!+\! k_{B} T\ \mathrm{log}\!\left(1\!-\!{\mathrm e}^{- \hbar \omega / k_{B} T}\right)\!\right]\!g(\vec{x}; \omega) d\omega,
% \nonumber
\end{equation}
where $g(\vec{x}; \omega)$ is the phonon density of states.
As mentioned before, calculation of the full phonon density of states is computationally demanding, requiring electronic structure calculations for multiple supercell configurations.
Instead, the ``GIBBS'' method uses a quasi-harmonic Debye model, where the Helmholtz free energy is expressed in terms of the Debye temperature $\theta_\Debye$
\begin{equation}
 \label{helmholtzdebye}
 A_{vib}(\theta_D; T) \!=\! n k_B T \!\left[ \frac{9}{8} \frac{\theta_\Debye}{T} \!+\! 3\ \mathrm{log}\!\left(1 \!-\! {\mathrm e}^{- \theta_\Debye / T}\!\right) \!\!-\!\! D\left(\frac{\theta_\Debye}{T}\right)\!\!\right],
% \nonumber
\end{equation}
where $n$ is the number of atoms in the unit cell and $D(\theta_\Debye / T)$ is the Debye integral
\begin{equation}
 D \left(\theta_\Debye/T \right) = 3 \left( \frac{T}{\theta_\Debye} \right)^3 \int_0^{\theta_\Debye/T} \frac{x^3}{e^x - 1} dx.
\end{equation}

In isotropic solids, 
changes in the geometry can be treated as isotropic changes in the volume, such that the magnitude of the configurational vector $\vec{x}$  is equal to the cube root of the volume, i.e. $x =
V^{\frac{1}{3}}$. 
The value of $\theta_D$ can thus be calculated as\cite{Blanco_CPC_GIBBS_2004,Blanco_jmolstrthch_1996,Poirier_Earth_Interior_2000}
\begin{equation}
 \label{debyetemp}
 \theta_D = \frac{\hbar}{k_B}[6 \pi^2 V^{1/2} n]^{1/3} f(\sigma) \sqrt{\frac{B_S}{M}}.
\end{equation}
Here, $M$ is the mass of the unit cell, $B_S$ is the adiabatic bulk modulus, and $f(\sigma)$ is given by
\begin{equation}
 \label{fpoisson}
 f(\sigma) = \left\{ 3 \left[ 2 \left( \frac{2}{3} \frac{1 + \sigma}{1 - 2 \sigma} \right)^{3/2} + \left( \frac{1}{3} \frac{1 + \sigma}{1 - \sigma} \right)^{3/2} \right]^{-1} \right\}^{\frac{1}{3}},
\end{equation}
in the assumption that the Poisson ratio $\sigma$ is constant.
The value of the Poisson ratio can be set as an input to \AGL\ separately from the DFT calculations, e.g., to the experimentally measured value.
For the calculations described in this paper this value is set at 0.25, which is the theoretical value for a Cauchy solid \cite{Blanco_CPC_GIBBS_2004, Poirier_Earth_Interior_2000}. 
The Poisson ratio $\sigma$ for crystalline materials is typically in the range of 0.2 to 0.3. 
Since the function $f(\sigma)$ behaves approximately linearly with values running from 0.9 to 0.7 when $\sigma$ is in the range from 0.2 to 0.3, this approximation has only a small effect on the results. 
We have checked this by performing the \AGL\ calculations using the literature values of the Poisson ratio where they are available. The correlation between calculated and
experimental values of the thermal conductivity is typically within a few percent of that obtained with the constant value of 0.25.

The adiabatic bulk modulus, B$_S$, can be approximated by the zero temperature limit  of the isothermal compressibility (neglecting zero-point contributions),  which we will refer to as B$_{\rm static}$:
\begin{eqnarray}
 \label{bulkmod}
 B_S &\approx& B_{\mathrm{static}} (\vec{x}) \approx B_{\mathrm{static}}(\vec{x}_{opt}(V)) =\\ \nonumber
 &=&V \left( \frac{\partial^2 E(\vec{x}_{opt} (V))}{\partial V^2} \right) = V \left( \frac{\partial^2 E(V)}{\partial V^2} \right),
\end{eqnarray}
where $\vec{x}_{opt}$ is the configuration vector of the unit cell geometry. 
The Gibbs free energy of the system can be expressed  as a function of the unit cell volume
\begin{equation}
 \label{gibbsdebye}
 G(V; p, T) = E(V) + pV + A_{\mathrm{vib}} (\theta_\Debye(V); T),
\end{equation}
where $\theta_D$ as a function of volume is evaluated from Equations (\ref{debyetemp}) and (\ref{bulkmod}), and $E(V)$ 
is obtained from a set of DFT calculations for unit cells with different volumes. 
Minimizing the Gibbs free energy with respect to volume, the equilibrium configuration at $(p, T)$ is determined, and additional properties, including the equilibrium 
$\theta_\Debye$, bulk modulus, heat capacity, thermal coefficient of expansion, etc. can be evaluated.

\subsection{Thermal calculation procedure}

In order to calculate the thermal properties for a particular material with a particular structure, first a set of DFT (e.g. {\small VASP} \cite{kresse_vasp}) calculations
which only differ by isotropic variations in the unit cell volume are set up and run using the {\small AFLOW} or Materials Project
framework. The resulting $E(V)$ is fitted by a polynomial, to calculate the adiabatic bulk modulus, $B_S$, as a function of
volume from Equation (\ref{bulkmod}). The $B_S$ values are then used to calculate the Debye temperature $\theta_\Debye$ for each unit cell
volume from Equation (\ref{debyetemp}). Next, the vibrational Helmholtz free energy $A_{vib}(\theta_\Debye (V); T)$ as a function of
volume, is calculated using Equation (\ref{helmholtzdebye}) for a given value of the temperature, $T$. The zero-pressure GIBBS free energy as a function of volume is then obtained by
\begin{equation}
 G(\vec{x}; 0, T) = E(V) + A_{\mathrm(vib)} (\theta_\Debye(V), T).
\end{equation}
This Gibbs free energy is fitted by a polynomial which is minimized with respect to
volume to find the equilibrium volume for any given value of $T$, at
zero pressure. 
The polynomial is an expansion in $x = \sqrt[3]{V}$. Therefore, finite 
pressure can be added simply to the coefficient of the $x^3$ term.
The volume which minimizes this modified polynomial for
$G(p, V, T)$ is the equilibrium volume that gives the Gibbs free energy for each requested $(p, T)$.
This equilibrium volume is used to calculate the bulk modulus and 
Debye
temperature of the material as a function of $p$ and $T$, from Equations
(\ref{bulkmod}) and (\ref{debyetemp}), respectively.

\subsection{DFT calculation details}

The DFT calculations to obtain $E(V)$ were performed using
the {\small VASP} software \cite{kresse_vasp} with projector- augmented-wave
pseudopotentials \cite{PAW} and the PBE parameterization of the
generalized gradient approximation to the exchange-correlation
functional \cite{PBE}. The energies were calculated at zero
temperature and pressure, 
with spin polarization and without zero-point motion or lattice
vibrations. The initial crystal structures were fully relaxed (cell
volume and shape and the basis atom coordinates inside the cell). 
An additional 100 different volume cells were calculated for each
structure by increasing or decreasing the relaxed lattice parameter in fractional
increments of 0.005. Numerical convergence to about 1 meV/atom was
ensured by a high-energy cut-off (30\% higher than the maximum cutoff
of each of the potentials) and a 8000 k-point Monkhorst-Pack
\cite{monkhorst} or $\Gamma$-centred (in the case of hexagonal unit
cells) mesh.

\subsection{The Gr{\"u}neisen Parameter}

The Gr{\"u}neisen parameter describes how the thermal properties of a material vary with the unit cell size.
It is defined by the phonon frequencies dependence on the unit cell volume
\begin{equation}
\label{gamma_micro}
 \gamma_i = - \frac{V}{\omega_i} \frac{\partial \omega_i}{\partial V}.
\end{equation}
Debye’s theory assumes that all mode frequencies
vary with the volume in the same ratio as the cut-off frequency
(Debye frequency), 
so the Gr{\"u}neisen parameter can be expressed in terms of $\theta_\Debye$
\begin{equation}
\label{gruneisen_theta}
 \gamma = - \frac{\partial \ \mathrm{log} (\theta_D(V))}{\partial \ \mathrm{log} V},
\end{equation}
and calculated using the Mie-Gr{\"u}neisen equation \cite{Poirier_Earth_Interior_2000}
\begin{equation}
\label{miegruneisen}
 p - p_{T=0} = \gamma \frac{U_{vib}}{V},
\end{equation}
where $U_{vib}$ is the vibrational internal energy
\begin{equation}
\label{Uvib}
 U_{vib} = n k_B T\left[ \frac{9}{8} \frac{\theta_D}{T} + 3D \left(
  \frac{\theta_D}{T} \right)\right].
\end{equation}

The expression in Eq. (\ref{gamma_micro}) can also be related to the
macroscopic definition via a weighted
average with the heat capacities for each branch of the phonon spectrum
\begin{equation}
 \gamma = \frac{\sum_i \gamma_i C_{V, i}} {\sum_i C_{V,i}}.
\end{equation}
that leads to the thermodynamic relations
\begin{equation}
 \gamma = V \left( \frac{\partial P}{\partial E} \right)_V = \frac{\alpha B_S}{C_P \rho} = \frac{\alpha B_T}{C_V \rho},
\end{equation}
where $\rho$ is the density of the material.

An alternative expression for the Gr{\"u}neisen parameter was derived
by Slater under the assumption of a constant Poisson ratio \cite{Slater_Chemical_Physics_1939}
\begin{equation}
\label{slatergruneisen}
\gamma = -\frac{2}{3} - \frac{V}{2} \frac{d^2p/dV^2}{dp/dV}.
\end{equation}

Equations (\ref{gruneisen_theta}), (\ref{miegruneisen}), and
(\ref {slatergruneisen}) have all been implemented within the \AGL.
Unless otherwise specified, the values of the Gr{\"u}neisen
parameter listed in the results and used to calculate the thermal
conductivity are obtained using Equation (\ref{miegruneisen}), as this
is generally considered more accurate than Equation
(\ref{gruneisen_theta}) \cite{Blanco_CPC_GIBBS_2004}.

\subsection{Thermal conductivity}

In the \AGL, the thermal conductivity is calculated by the method
proposed by Slack \cite{slack, Morelli_Slack_2006} using the Debye temperature and the
Gr{\"u}neisen parameter 
\begin{eqnarray}
 \label{thermal_conductivity}
 \kappa_l (\theta_\acoustic) &=& \frac{0.849 \times 3 \sqrt[3]{4}}{20 \pi^3(1 - 0.514\gamma^{-1} + 0.228\gamma^{-2})} \times \\ \nonumber
 & &\times \left( \frac{k_B \theta_\acoustic}{\hbar} \right)^2 \frac{k_B m V^{\frac{1}{3}}}{\hbar \gamma^2}.
\end{eqnarray}
where $V$ is the volume of the unit cell and $m$ is the average atomic
mass.
It should be noted that the Debye
temperature in this formula, $\theta_\acoustic$, is slightly
different than the traditional
Debye temperature, $\theta_\Debye$, calculated in Equation
(\ref{debyetemp}). Instead, $\theta_\acoustic$ is obtained by only
considering the acoustic modes, 
based on the assumption that the optical
phonon modes in crystals do not contribute to heat transport \cite{slack}.
This $\theta_\acoustic$ is referred to as the ``acoustic'' Debye temperature
\cite{slack, Morelli_Slack_2006}. It can be derived directly from
the phonon DOS by integrating only over the acoustic modes \cite{slack,
  Wee_Fornari_TiNiSn_JEM_2012}. Alternatively, it 
can be calculated from the traditional Debye temperature $\theta_\Debye$ \cite{slack, Morelli_Slack_2006}
\begin{equation}
\label{acousticdebyetemp}
\theta_\acoustic = \theta_\Debye n^{-\frac{1}{3}}.
\end{equation}
To demonstrate the distinction between these two quantities, we
include the values of both $\theta_\Debye$ and $\theta_\acoustic$, as
calculated using \AGL, in the
tables of results in the following sections.

The thermal conductivity at temperatures other than $\theta_\acoustic$ is estimated by \cite{slack, Morelli_Slack_2006, Madsen_PRB_2014}:
\begin{equation}
 \label{kappa_temperature}
 \kappa_l(T) = \kappa_l(\theta_\acoustic) \frac{\theta_\acoustic}{T}.
\end{equation}

In principle, the Gr{\"u}neisen
parameter in Equation (\ref{thermal_conductivity}) should also be
derived only from the
acoustic phonon modes \cite{slack}. However, unlike the case of
$\theta_\Debye$ and $\theta_\acoustic$, 
there is no simple way to extract
it from the traditional Gr{\"u}neisen parameter. Instead, it must be
calculated from Equation (\ref{gamma_micro})
for each phonon branch
separately and summed over
the acoustic branches.
This requires calculating the full phonon spectrum for different
volumes, and is therefore too computationally demanding to be used for
high-throughput screening. The dependence of the expression 
(\ref{thermal_conductivity}) on $\gamma$ is weak \cite{Morelli_Slack_2006}, thus 
the evaluation of $\kappa_l$ using the traditional  Gr{\"u}neisen parameter
 introduces just a small systematic error which is insignificant for
 screening purposes. 

\subsection{Pearson and Spearman Correlations}

Pearson and Spearman correlations have been implemented separately
from \AGL, in order to
analyze the results for entire sets of materials. 
The Pearson correlation coefficient $r$ is a measure of the linear
correlation between two variables, $X$ and $Y$.
It is calculated by 
\begin{equation} 
 r = \frac{\sum_{i=1}^{n} \left(X_i - \overline{X} \right) \left(Y_i - \overline{Y} \right) }{ \sqrt{\sum_{i=1}^{n} \left(X_i - \overline{X} \right)^2} \sqrt{\sum_{i=1}^{n} \left(Y_i - \overline{Y} \right)^2}},
\end{equation}
where $\overline{X}$ and $\overline{Y}$ are the mean values of $X$ and $Y$.

The Spearman rank correlation coefficient $\rho$ is a measure of the
monotonicity of the relation between two variables. 
The raw values of the two variables $X_i$ and $Y_i$ are sorted in
ascending order, and are assigned rank values $x_i$ and $y_i$ which
are equal to their position in the sorted list. If there is more than
one variable with the same value, the average of the position values
are assigned to each. The correlation
coefficient is then given by
\begin{equation} 
 \rho = \frac{\sum_{i=1}^{n} \left(x_i - \overline{x} \right)
 \left(y_i - \overline{y} \right) }{ \sqrt{\sum_{i=1}^{n} \left(x_i
 - \overline{x} \right)^2} \sqrt{\sum_{i=1}^{n} \left(y_i -
 \overline{y} \right)^2}}.
\end{equation}

It is useful for determining how well the ranking order of the values
of one variable predict the ranking order of the values of the other variable.

\section{Results}

We used the  \AGL\ to calculate the the Debye
temperature, Gr{\"u}neisen parameter and thermal conductivity for
a set of 75 materials
with the diamond, zincblende, rocksalt and wurzite structures, and 107
half-Heusler compounds. 
The results have been compared to 
first-principles calculations (and experimental values where available) of the half-Heusler compounds and to
experimental values for the other structures. 

\subsection{Zincblende and Diamond structure materials}

\begin{figure}[t!]
 \includegraphics[width=0.98\columnwidth]{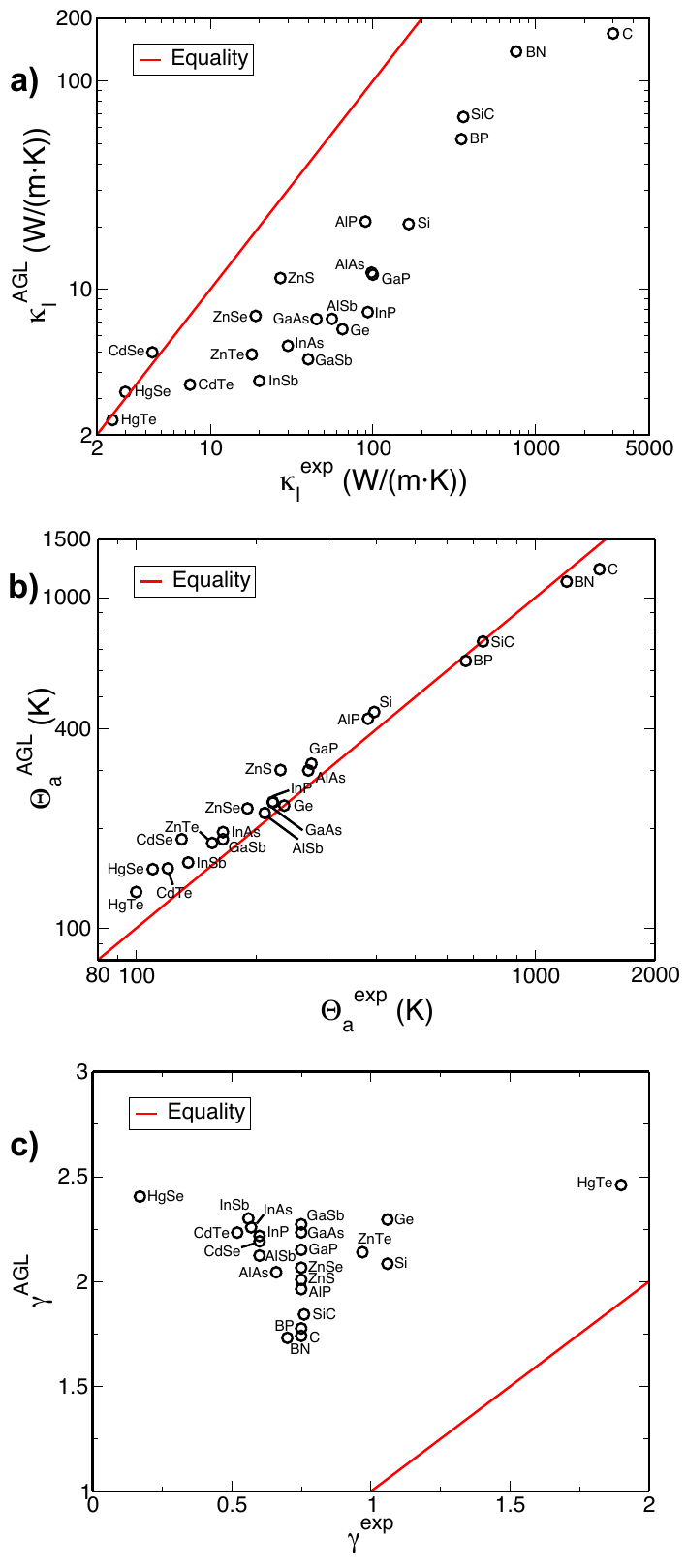}
 \vspace{-4mm}
 \caption{\small {\bf (a)} Lattice thermal conductivity at 300K, {\bf
     (b)} acoustic Debye temperature
  and {\bf (c)} Gr{\"u}neisen parameter of Zincblende and Diamond structure semiconductors.
 \label{fig:zincblende}
}
\end{figure}

Experimental values of thermal properties for materials with the
Zincblende (spacegroup: $F\overline{4}3m\ (\#216)$; Pearson symbol: $cF8$) and
Diamond (spacegroup: $Fd\overline{3}m\ (\#227)$; Pearson symbol: $cF8$) structures were published in Table
II of Ref. \onlinecite{slack} and Table 2.2 of
Ref. \onlinecite{Morelli_Slack_2006}.
They are shown with the calculated thermal conductivity at 300K, the Debye temperature
and Gr{\"u}neisen parameter for these materials in Table \ref{tab:zincblende} and in Figure
\ref{fig:zincblende}. As shown in the table, for a few of these
materials there are discrepancies in
experimental values quoted in the different sources. 
For each entry we
used the value from the most recent source 
for plotting the following figures and calculating the
correlations reported here.

\begin{table}[t!]
 \caption{\small Lattice thermal conductivity at 300K, Debye
   temperature and Gr{\"u}neisen parameter of Zincblende and Diamond
   structure semiconductors. The values 
   listed for $\theta^\EXP$ are 
    $\theta_\acoustic$, except 141K for HgTe which is $\theta_\Debye$ \cite{Snyder_jmatchem_2011}.
  Units: $\theta$ in \K, $\kappa$ in \WmK.
 }
 \label{tab:zincblende}
 {\footnotesize
 \begin{tabular}{c c c c c c c c}
  \hline
  Comp. & $\theta^\EXP$  & $\theta_\acoustic^\AGL$ & $\theta_\Debye^\AGL$ & $\gamma^\EXP$ & $\gamma^\AGL$ & $\kappa^\EXP$  & $\kappa^\AGL$    \\
  %  &  \K &   \K &   &  &  \WmK  & \WmK   \\
%    Compound & $\theta_\Debye^\EXP \K$  & $\theta_\Debye^\AGL$ \K & $\gamma^\EXP$ & $\gamma^\AGL$ & $\kappa^\EXP$   \WmK  & $\kappa^\AGL$   \WmK    \\
     \hline
    C & 1450 \cite{slack, Morelli_Slack_2006} & 1219 & 1536 & 0.75 \cite{Morelli_Slack_2006} & 1.74 &  3000 \cite{Morelli_Slack_2006} & 169.1 \\
     & & &  & 0.9 \cite{slack} & &  & \\
    SiC & 740 \cite{slack} & 737 & 928 & 0.76 \cite{slack} & 1.84 & 360 \cite{Ioffe_Inst_DB} & 67.19 \\
    Si & 395 \cite{slack, Morelli_Slack_2006} & 451 & 568 & 1.06 \cite{Morelli_Slack_2006} & 2.09 & 166 \cite{Morelli_Slack_2006} & 20.58 \\
     & &  &  & 0.56 \cite{slack} &  &  & \\
    Ge & 235 \cite{slack, Morelli_Slack_2006} & 235 & 296 & 1.06 \cite{Morelli_Slack_2006} & 2.3 &  65 \cite{Morelli_Slack_2006} &  6.44  \\
    & & & & 0.76 \cite{slack} & &  &  \\
    BN & 1200 \cite{Morelli_Slack_2006} & 1118 & 1409 & 0.7 \cite{Morelli_Slack_2006} & 1.73 & 760 \cite{Morelli_Slack_2006} & 138.38 \\
    BP & 670 \cite{slack, Morelli_Slack_2006} & 644 & 811 & 0.75 \cite{Morelli_Slack_2006} & 1.78 & 350 \cite{Morelli_Slack_2006} & 52.56 \\
    % BAs & 404 \cite{Morelli_Slack_2006} & 529 & 0.75 \cite{Morelli_Slack_2006} & 1.95 & 2200 \cite{Lindsay_PRL_2013} & 51.5  \\
    AlP & 381 \cite{Morelli_Slack_2006} & 430 & 542 & 0.75 \cite{Morelli_Slack_2006} & 1.96 & 90 \cite{Landolt-Bornstein, Spitzer_JPCS_1970} & 21.16 \\
    AlAs & 270 \cite{slack, Morelli_Slack_2006} & 300 & 378 & 0.66 \cite{slack, Morelli_Slack_2006} & 2.04 & 98 \cite{Morelli_Slack_2006} &  12.03 \\
    AlSb & 210 \cite{slack, Morelli_Slack_2006} & 223 & 281 & 0.6 \cite{slack, Morelli_Slack_2006} & 2.12 & 56 \cite{Morelli_Slack_2006} & 7.22 \\
    GaP & 275 \cite{slack, Morelli_Slack_2006} & 314 & 396 & 0.75 \cite{Morelli_Slack_2006} & 2.15 & 100  \cite{Morelli_Slack_2006} & 11.76 \\
     &  & &  & 0.76 \cite{slack} &  &  &  \\
    GaAs & 220 \cite{slack, Morelli_Slack_2006} & 240 & 302 & 0.75 \cite{slack, Morelli_Slack_2006} & 2.23 & 45 \cite{Morelli_Slack_2006} & 7.2 \\
    GaSb & 165 \cite{slack, Morelli_Slack_2006} & 186 & 234 & 0.75 \cite{slack, Morelli_Slack_2006} & 2.27 & 40 \cite{Morelli_Slack_2006} & 4.62 \\
    InP  & 220 \cite{slack, Morelli_Slack_2006} & 241 & 304 & 0.6 \cite{slack, Morelli_Slack_2006} & 2.22 & 93 \cite{Morelli_Slack_2006} & 7.78  \\
    InAs & 165 \cite{slack, Morelli_Slack_2006} & 195 &  246 & 0.57 \cite{slack, Morelli_Slack_2006} & 2.26 & 30 \cite{Morelli_Slack_2006} & 5.36 \\
    InSb & 135 \cite{slack, Morelli_Slack_2006} & 158 & 199 & 0.56 \cite{slack, Morelli_Slack_2006} & 2.3 & 20 \cite{Morelli_Slack_2006} & 3.64 \\
     & & &  & &  & 16.5 \cite{Snyder_jmatchem_2011} &  \\
    ZnS & 230 \cite{slack, Morelli_Slack_2006} & 301 & 379 & 0.75 \cite{slack, Morelli_Slack_2006} & 2.01 & 27 \cite{Morelli_Slack_2006} & 11.33 \\
    ZnSe & 190 \cite{slack, Morelli_Slack_2006} & 230 & 290 & 0.75 \cite{slack, Morelli_Slack_2006} & 2.07 & 19 \cite{Morelli_Slack_2006} & 7.46 \\
     &  & & &  &  & 33  \cite{Snyder_jmatchem_2011} &  \\
    ZnTe & 155 \cite{slack, Morelli_Slack_2006} & 181 & 228 & 0.97 \cite{slack, Morelli_Slack_2006} & 2.14 & 18 \cite{Morelli_Slack_2006} &  4.87  \\
    CdSe & 130 \cite{Morelli_Slack_2006} & 186 & 234 & 0.6 \cite{Morelli_Slack_2006} & 2.19 & 4.4 \cite{Snyder_jmatchem_2011} & 4.99 \\
    CdTe & 120 \cite{slack, Morelli_Slack_2006} & 152 & 191 & 0.52 \cite{slack, Morelli_Slack_2006} & 2.23 & 7.5 \cite{Morelli_Slack_2006} & 3.49 \\
    HgSe & 110 \cite{slack} & 151 & 190 & 0.17 \cite{slack} & 2.4 & 3 \cite{Whitsett_PRB_1973} & 3.22 \\
    HgTe & 141 \cite{Snyder_jmatchem_2011} & 129 & 162 & 1.9 \cite{Snyder_jmatchem_2011}  & 2.46 & 2.5 \cite{Snyder_jmatchem_2011}  & 2.36 \\
     & 100 \cite{slack} & & & 0.46\cite{slack}  & &  &  \\

    \hline
  \end{tabular}
}
\end{table}

Comparison of the calculated and experimental results of Table
\ref{tab:zincblende} shows that the absolute agreement between them is
quite poor, with discrepancies of tens, or even hundreds, of percent
quite common. Considerable disagreements also
exist between different experimental reports of these properties, in
almost all cases where they exist. Unfortunately, the scarcity of experimental data
from different sources on the thermal properties of these materials
prevents reaching definite conclusions regarding the true values of
these properties. The available data can thus only be considered as a rough indication of their
order of magnitude.

Nevertheless, the Pearson correlation between the
\AGL\ calculated thermal conductivity values and the experimental
values is high, $0.878$. The Spearman correlation, $0.905$, is even
higher. The
Spearman correlation between the experimental values of the thermal
conductivity and $\theta_\acoustic$ as calculated with \AGL\
is $0.925$.
There is also a strong correlation between the experimental values of $\theta_\acoustic$
and those calculated with \AGL, with a
Pearson correlation of $0.995$ and a Spearman correlation of
$0.984$. 
The correlation for the Gr{\"u}neisen parameter is
much worse, with Pearson and Spearman correlations of $0.137$ and
$-0.187$, respectively.

Table 2.2 of Ref. \onlinecite{Morelli_Slack_2006} includes 
values of the thermal conductivity at 300K, calculated using the
experimental values of $\theta_\acoustic$ and $\gamma$. The Pearson correlation between these
calculated thermal conductivity values and the experimental
values is $0.932$, and the corresponding Spearman correlation is
$0.941$. Both values are just slightly higher than the correlations we
calculated using the \AGL\ evaluations of $\theta_\acoustic$ and
$\gamma$. Thus, the unsatisfactory quantitative
reproduction of these quantities by the Debye quasi-harmonic model
has little impact on its effectiveness as a screening tool for high or
low thermal conductivity materials. The model can be used when these
experimental values are unavailable.

These results indicate that despite the quantitative disagreement
between the calculated and experimental results for the thermal
conductivity and $\theta_\acoustic$, the \AGL\ calculations are
good indicators for the relative values of these quantities and for
ranking materials in order of increasing conductivity. For the Diamond
and Zincblende structure materials, the calculated $\theta_\acoustic$ turns out to be a slightly better indicator of the ordinal order of
the thermal conductivity than the calculated conductivity.

\subsection{Rocksalt structure materials}

\begin{figure}[t!]
  \includegraphics[width=0.98\columnwidth]{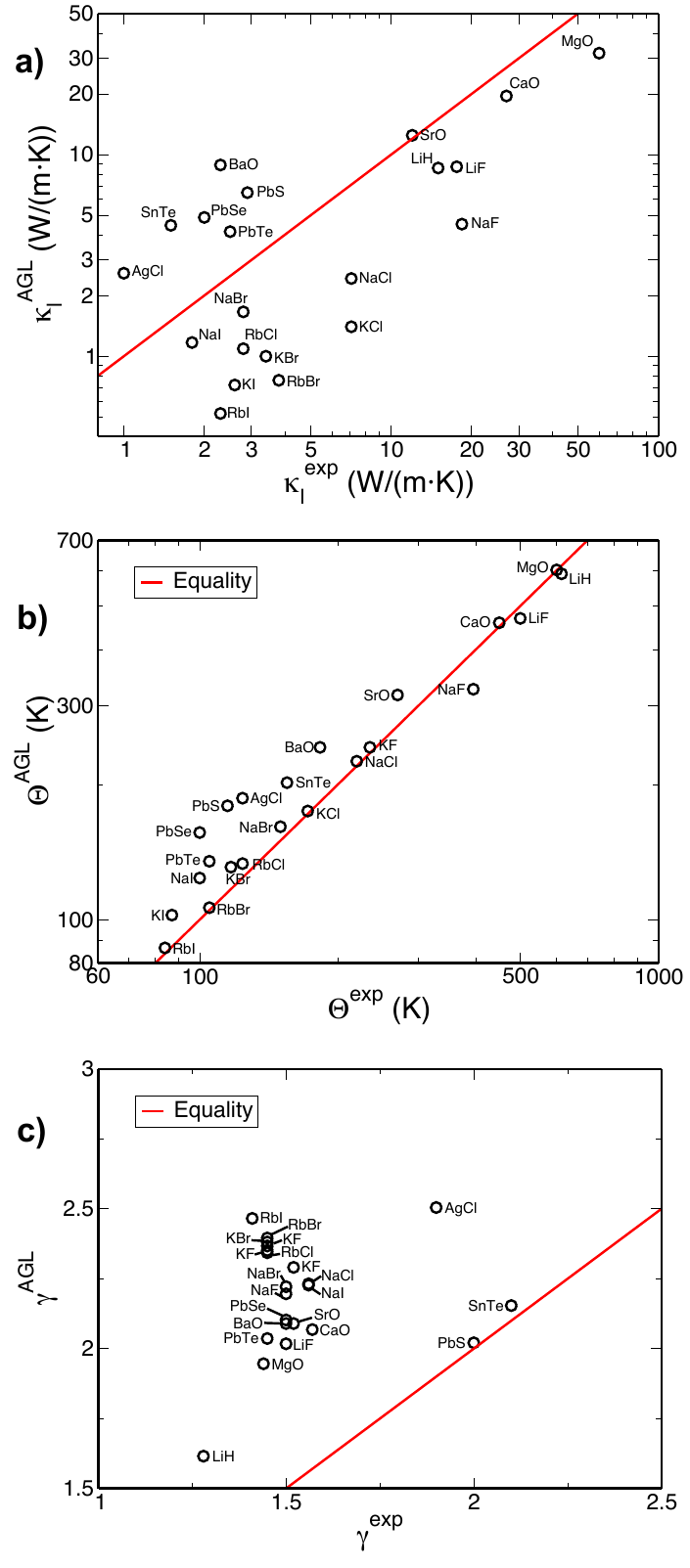}
  \vspace{-4mm}
  \caption{\small  {\bf (a)} Lattice thermal conductivity at 300K,
    {\bf (b)} Debye temperature
    and {\bf (c)} Gr{\"u}neisen parameter of Rocksalt structure
    semiconductors. 
The Debye temperatures plotted in {\bf (b)} are
    $\theta_\acoustic$, except for SnTe where $\theta_\Debye$ is
    quoted in Ref. \onlinecite{Snyder_jmatchem_2011}.}
  \label{fig:rocksalt}
\end{figure}

Experimental values of the thermal properties of materials with the
Rocksalt structure (spacegroup: $Fm\overline{3}m\ (\#225)$; Pearson symbol: $cF8$) 
were published in Table III of
Ref. \onlinecite{slack} and Table 2.1 of
Ref. \onlinecite{Morelli_Slack_2006}.
They are compared to the values calculated by the \AGL\ 
in Table \ref{tab:rocksalt} and  Figure   \ref{fig:rocksalt}.
As was the case for the zincblende structure materials, we have
included the \AGL\ results for both $\theta_\Debye$ and
$\theta_\acoustic$ in the table.
The experimental values listed in the table are all for $\theta_\acoustic$
\cite{Morelli_Slack_2006}, with the exception of the value of 155K for
SnTe, which is for $\theta_\Debye$\cite{Snyder_jmatchem_2011}. The \AGL\
$\theta_\acoustic$ values were used for plotting and correlation
calculations, with the exception of that for SnTe where
$\theta_\Debye$ was used for plotting Figure \ref{fig:rocksalt}b and
for calculating the correlation between the Debye temperatures.

The Pearson correlation between the calculated and experimental values
for the thermal conductivity 
is $0.910$. The Spearman correlation is $0.445$. The Spearman
correlation between the experimental
values of the thermal conductivity
and the calculated values of $\theta_\acoustic$ is $0.645$.
The Pearson correlation between the calculated and experimental values
for the Debye temperature is $0.982$ and the corresponding Spearman correlation is
$0.948$. 
 The correlation for the Gr{\"u}neisen parameter is
much worse, with Pearson and Spearman correlations of $0.118$ and
$-0.064$, respectively.

\begin{table}[t!]
  \caption{\small Lattice thermal conductivity, Debye temperature
    and Gr{\"u}neisen parameter of Rocksalt structure semiconductors. 
 The values listed for $\theta^{exp}$ are
    $\theta_\acoustic$, except 155K for SnTe which is $\theta_\Debye$ \cite{Snyder_jmatchem_2011}.
   Units: $\theta$ in \K, $\kappa$ in \WmK.
 }
  \label{tab:rocksalt}
  \begin{tabular}{c c c c c c c c}
    \hline
    Comp. & $\theta^\EXP$  & $\theta_\acoustic^\AGL$ & $\theta_\Debye^\AGL$ & $\gamma^\EXP$ & $\gamma^\AGL$ & $\kappa^\EXP$  & $\kappa^\AGL$    \\
 %  &  \K &   \K &   &  &  \WmK  & \WmK   \\
 %   Compound & $\theta_\Debye^\EXP \K$  & $\theta_\Debye^\AGL$ \K & $\gamma^\EXP$ & $\gamma^\AGL$ & $\kappa^\EXP$   \WmK  & $\kappa^\AGL$   \WmK    \\
     \hline
    LiH & 615 \cite{slack, Morelli_Slack_2006} & 590 & 743 & 1.28 \cite{slack, Morelli_Slack_2006} & 1.62 & 15 \cite{Morelli_Slack_2006} & 8.58 \\
    LiF & 500 \cite{slack, Morelli_Slack_2006} & 469 & 591 & 1.5 \cite{slack, Morelli_Slack_2006} & 2.02 & 17.6 \cite{Morelli_Slack_2006} & 8.71 \\
    NaF & 395 \cite{slack, Morelli_Slack_2006} & 326 & 411 & 1.5 \cite{slack, Morelli_Slack_2006} & 2.2 &  18.4 \cite{Morelli_Slack_2006} &  4.52  \\
    NaCl & 220 \cite{slack, Morelli_Slack_2006} & 225 & 284 & 1.56 \cite{slack, Morelli_Slack_2006} & 2.23 & 7.1 \cite{Morelli_Slack_2006} & 2.43 \\
    NaBr & 150 \cite{slack, Morelli_Slack_2006} & 161 & 203 & 1.5 \cite{slack, Morelli_Slack_2006} & 2.22 & 2.8 \cite{Morelli_Slack_2006} & 1.66 \\
    NaI & 100 \cite{slack, Morelli_Slack_2006} & 124 & 156 & 1.56 \cite{slack, Morelli_Slack_2006} & 2.23 & 1.8 \cite{Morelli_Slack_2006} & 1.17 \\
    KF & 235 \cite{slack, Morelli_Slack_2006} & 242 & 305 & 1.52 \cite{slack, Morelli_Slack_2006} & 2.29 &  &  2.68 \\
    KCl & 172 \cite{slack, Morelli_Slack_2006} & 175 & 220 & 1.45 \cite{slack, Morelli_Slack_2006} & 2.38 & 7.1 \cite{Morelli_Slack_2006} & 1.4 \\
    KBr & 117 \cite{slack, Morelli_Slack_2006} & 131 & 165 & 1.45 \cite{slack, Morelli_Slack_2006} & 2.37 & 3.4 \cite{Morelli_Slack_2006} &  1.0 \\
    KI & 87 \cite{slack, Morelli_Slack_2006} & 102 & 129 & 1.45 \cite{slack, Morelli_Slack_2006} & 2.35 & 2.6 \cite{Morelli_Slack_2006} & 0.72 \\
    RbCl & 124 \cite{slack, Morelli_Slack_2006} & 133 & 168 & 1.45 \cite{slack, Morelli_Slack_2006} & 2.34 & 2.8 \cite{Morelli_Slack_2006} & 1.09 \\
    RbBr & 105 \cite{slack, Morelli_Slack_2006} & 106 & 134 & 1.45 \cite{slack, Morelli_Slack_2006} & 2.40 & 3.8 \cite{Morelli_Slack_2006} & 0.76 \\
    RbI & 84 \cite{slack, Morelli_Slack_2006} & 87 & 109 & 1.41 \cite{slack, Morelli_Slack_2006} & 2.47 & 2.3 \cite{Morelli_Slack_2006} & 0.52 \\
    AgCl & 124 \cite{slack} & 187 & 235 & 1.9 \cite{slack} & 2.5 & 1.0 \cite{Landolt-Bornstein, Maqsood_IJT_2003}  & 2.58 \\
    MgO  & 600 \cite{slack, Morelli_Slack_2006} & 602 & 758 & 1.44 \cite{slack, Morelli_Slack_2006} & 1.95 & 60 \cite{Morelli_Slack_2006} & 31.86  \\
    CaO & 450 \cite{slack, Morelli_Slack_2006} & 459 & 578 & 1.57 \cite{slack, Morelli_Slack_2006} & 2.07 & 27 \cite{Morelli_Slack_2006} & 19.54 \\
    SrO & 270 \cite{slack, Morelli_Slack_2006} & 317 & 399 & 1.52 \cite{slack, Morelli_Slack_2006} & 2.09 & 12 \cite{Morelli_Slack_2006} & 12.47 \\
    BaO & 183 \cite{slack, Morelli_Slack_2006} & 242 & 305 & 1.5 \cite{slack, Morelli_Slack_2006} & 2.09 & 2.3 \cite{Morelli_Slack_2006} & 8.88 \\
    PbS & 115 \cite{slack, Morelli_Slack_2006} & 179 & 226 & 2.0 \cite{slack, Morelli_Slack_2006} & 2.02 & 2.9 \cite{Morelli_Slack_2006} & 6.48 \\
    PbSe & 100 \cite{Morelli_Slack_2006} & 156 & 197 & 1.5 \cite{Morelli_Slack_2006} & 2.1 & 2.0 \cite{Morelli_Slack_2006} & 4.88 \\
    PbTe & 105 \cite{slack, Morelli_Slack_2006} & 135 & 170 & 1.45 \cite{slack, Morelli_Slack_2006} & 2.04 & 2.5 \cite{Morelli_Slack_2006} & 4.15 \\
    SnTe & 155 \cite{Snyder_jmatchem_2011} & 160 & 202 & 2.1 \cite{Snyder_jmatchem_2011} & 2.15 & 1.5 \cite{Snyder_jmatchem_2011} & 4.46 \\
    \hline
  \end{tabular}
\end{table}

Table 2.1 of Ref. \onlinecite{Morelli_Slack_2006} includes 
values of the thermal conductivity at 300K, calculated using the
experimental values of $\theta_\acoustic$ and $\gamma$. The Pearson correlation between these
calculated thermal conductivities and their experimental
values is $0.986$, and the corresponding Spearman correlation is
$0.761$.  Comparing these values with the correlations obtained using
the \AGL\ calculated quantities, we find that the latter
 are more significantly degraded than for the Diamond and Zincblende structures.
This is despite the similar correlations obtained for $\theta_\acoustic$
and $\gamma$ in these two cases.
Nevertheless, the Pearson correlation between the calculated and
experimental conductivities is high in both calculations, indicating that the \AGL\
approach may be used as a screening tool for high conductivity
compounds in cases where gaps exist in the experimental data for these
materials.

\subsection{Wurzite structure materials}

\begin{table}[t!]
  \caption{\small Lattice thermal conductivity at 300K, Debye temperature and
    Gr{\"u}neisen parameter of Wurzite structure semiconductors. 
    The experimental Debye temperature values listed are
    $\theta_\acoustic$, except 190K for InSe
    \cite{Snyder_jmatchem_2011} and 660K for InN \cite{Ioffe_Inst_DB,
      Krukowski_jphyschemsolids_1998} which are $\theta_\Debye$.
   Units: $\theta$ in \K, $\kappa$ in \WmK.
}
  \label{tab:wurzite}
  \begin{tabular}{c c c c c c c c}
    \hline
    Comp. & $\theta^\EXP$  & $\theta_\acoustic^\AGL$ & $\theta_\Debye^\AGL$ & $\gamma^\EXP$ & $\gamma^\AGL$ & $\kappa^\EXP$  & $\kappa^\AGL$    \\
 %   &  \K &   \K &   &  &  \WmK  & \WmK   \\
 %   Compound & $\theta_\Debye^\EXP \K$  & $\theta_\Debye^\AGL$ \K & $\gamma^\EXP$ & $\gamma^\AGL$ & $\kappa^\EXP$   \WmK  & $\kappa^\AGL$   \WmK    \\
     \hline
    SiC & 740 \cite{Morelli_Slack_2006} & 750 & 1191 & 0.75 \cite{Morelli_Slack_2006} & 1.86 & 490 \cite{Morelli_Slack_2006} & 52.63 \\
    AlN & 620 \cite{Morelli_Slack_2006} & 485 & 770 & 0.7 \cite{Morelli_Slack_2006} & 1.85 & 350 \cite{Morelli_Slack_2006} & 32.58 \\
    GaN & 390 \cite{Morelli_Slack_2006} & 291 & 462 & 0.7 \cite{Morelli_Slack_2006} & 2.07 &  210 \cite{Morelli_Slack_2006} &  14.55  \\
    ZnO & 303 \cite{Morelli_Slack_2006} & 519 & 824 & 0.75 \cite{Morelli_Slack_2006} & 1.97 & 60 \cite{Morelli_Slack_2006} & 20.98 \\
    BeO & 809 \cite{Morelli_Slack_2006} & 784 & 1244 & 1.38 \cite{Slack_JAP_1975, Cline_JAP_1967, Morelli_Slack_2006} & 1.76 & 370 \cite{Morelli_Slack_2006} & 44.6 \\
     & & & & 0.75 \cite{Morelli_Slack_2006} &  & & \\
    CdS & 135 \cite{Morelli_Slack_2006} & 146 & 231 & 0.75 \cite{Morelli_Slack_2006} & 2.14 & 16 \cite{Morelli_Slack_2006} & 3.59 \\
    InSe & 190 \cite{Snyder_jmatchem_2011} & 106 & 212 & 1.2 \cite{Snyder_jmatchem_2011} & 2.24 & 6.9 \cite{Snyder_jmatchem_2011} &  1.72 \\
    InN & 660 \cite{Ioffe_Inst_DB, Krukowski_jphyschemsolids_1998} & 202 & 321 & 0.97 \cite{Krukowski_jphyschemsolids_1998} & 2.17 & 45 \cite{Ioffe_Inst_DB, Krukowski_jphyschemsolids_1998} & 8.04 \\
    \hline
  \end{tabular}
\end{table}

\begin{figure}[t!]
  \includegraphics[width=0.98\columnwidth]{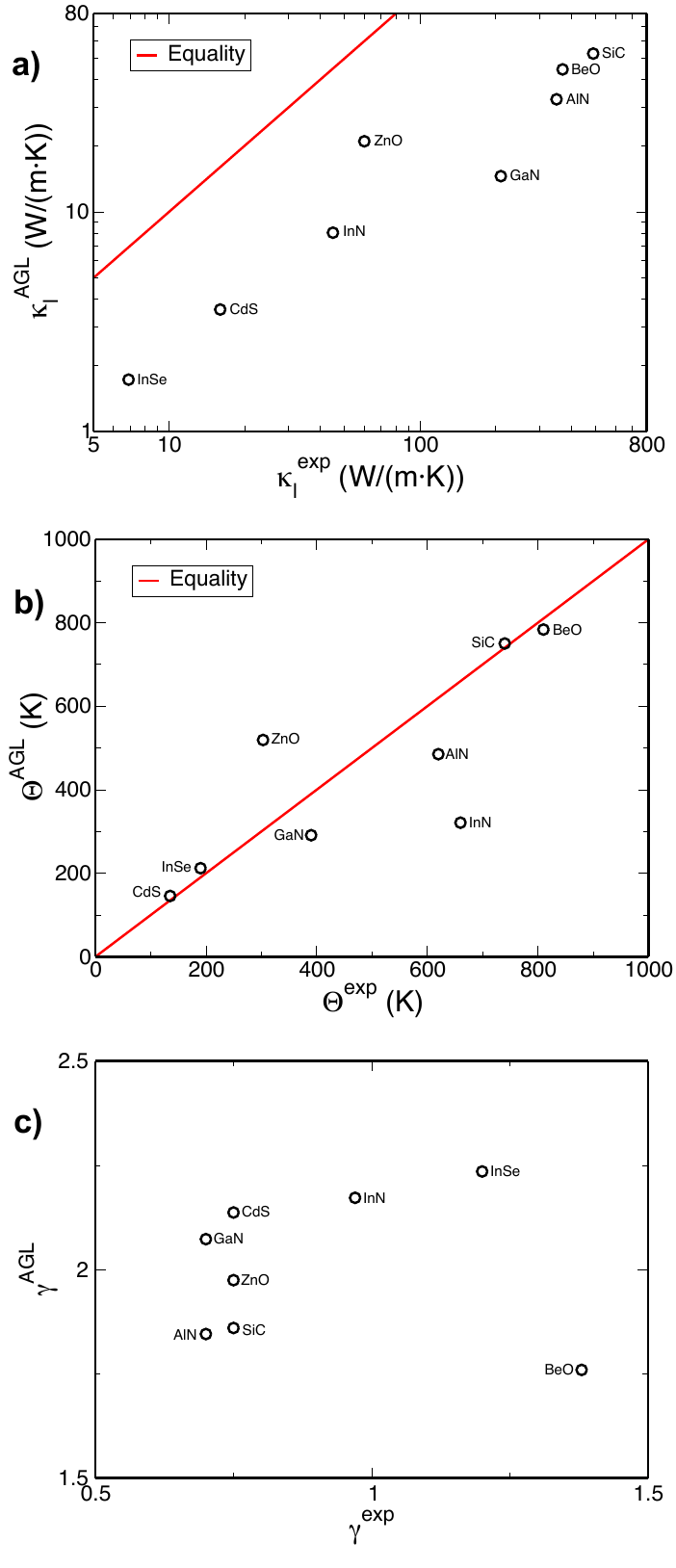}
  \vspace{-4mm}
  \caption{\small  {\bf (a)} Lattice thermal conductivity at 300K,
    {\bf (b)} Debye temperature
    and {\bf (c)} Gr{\"u}neisen parameter of Wurzite structure
    semiconductors.
 The Debye temperatures plotted in {\bf (b)} are
    $\theta_\acoustic$, except for InSe and InN where $\theta_\Debye$ 
     values are quoted in Refs.
    \onlinecite{Snyder_jmatchem_2011, Ioffe_Inst_DB, Krukowski_jphyschemsolids_1998}.}
  \label{fig:wurzite}
\end{figure}

Experimental results for Wurzite structure materials (spacegroup:
$P6_3mc\ (\#186)$; Pearson symbol: $hP4$) appear
in Table 2.3 of Ref. \onlinecite{Morelli_Slack_2006}.
Their comparison with our calculation results is shown in Table
\ref{tab:wurzite} and Figure   \ref{fig:wurzite}. 
As was the case for the zincblende and wurzite structure materials, we have
included the \AGL\ results for both $\theta_\Debye$ and $\theta_\acoustic$ in the table,
while the \AGL\ $\theta_\acoustic$ was used for plotting Figure
\ref{fig:wurzite} and calculating the correlations.
The experimental values listed in the table are all for $\theta_\acoustic$
\cite{Morelli_Slack_2006}, with the exceptions of the values of 190K for InSe \cite{Snyder_jmatchem_2011} and 660K
for InN \cite{Ioffe_Inst_DB, Krukowski_jphyschemsolids_1998}, which
are for $\theta_\Debye$. The \AGL\
$\theta_\acoustic$ values were used for plotting and correlation
calculations, with the exception of that those for InSe and InN where
$\theta_\Debye$ was used for plotting Figure \ref{fig:wurzite}b and
for calculating the correlation between the Debye temperatures.

The Pearson correlation between the \AGL\ thermal conductivity
values and the experimental values
is $0.943$. The corresponding Spearman correlation
is $0.976$. The Spearman correlation between the experimental values
of the thermal conductivity and
the calculated values of $\theta_\acoustic$ is $0.905$.
The Pearson correlation between the experimental and calculated
values of the Debye temperature is $0.8$, and the corresponding Spearman correlation is
$0.833$. 
The correlations for the Gr{\"u}neisen parameter are both
poor, with Pearson and Spearman values of $-0.039$ and
$0.160$, respectively.

Table 2.3 of Ref. \onlinecite{Morelli_Slack_2006} includes 
values of the thermal conductivity at 300K, calculated using the
experimental values of the Debye temperature and Gr{\"u}neisen parameter. The Pearson correlation between these
calculated thermal conductivity values and the experimental
values is $0.996$, and the corresponding Spearman correlation is
$1.0$.  These values are again higher than the correlations obtained using
the \AGL\ calculated quantities, however, all of these
correlations are very high so either of the calculation methods
could serve as a reliable screening tool of the thermal conductivity.
It should be noted that the high correlations calculated with the
experimental $\theta_\acoustic$ and $\gamma$ were obtained using
$\gamma=0.75$ for BeO. Table 2.3 of
Ref. \onlinecite{Morelli_Slack_2006} also cites an alternative value
of $\gamma=1.38$ for BeO (Table \ref{tab:wurzite}). Using this outlier
value would severely degrade the results down to $0.7$, for the
Pearson correlation, and $0.829$, for the Spearman correlation.
These values are too low for a reliable screening tool. This
demonstrates the ability of the
\AGL\ calculations to compensate for anomalies in the
experimental data when
they exist and still provide a reliable screening method for the
thermal conductivity.

\subsection{Rhombohedral materials}

\begin{figure}[t!]
  \includegraphics[width=0.98\columnwidth]{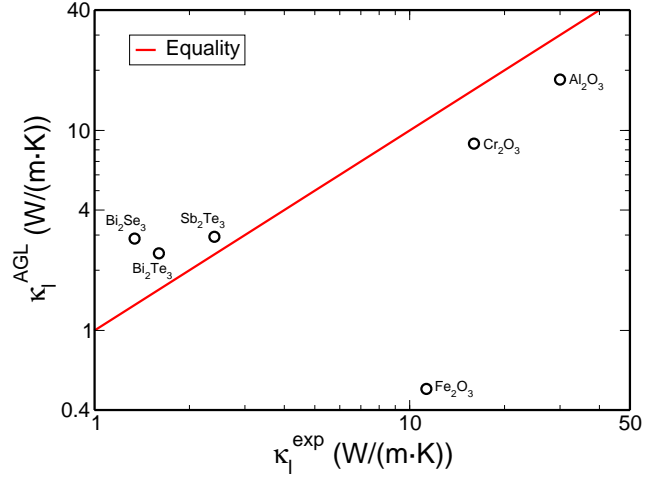}
  \vspace{-4mm}
  \caption{\small Lattice thermal conductivity of rhombohedral
    semiconductors at 300K.}
  \label{fig:rhombo}
\end{figure}

Experimental results for rhombohedral materials (spacegroups:
$R\overline{3}mR\ (\#166)$, $R\overline{3}mH\ (\#166)$
and $R\overline{3}cH\ (\#167)$; Pearson symbols: $hR5$, $hR10$)
are compared to the results of our
calculations in Table \ref{tab:rhombo}
and Figure  \ref{fig:rhombo}.
The experimental Debye temperatures are for $\theta_\Debye$ in the
case of Bi$_2$Te$_3$ and Sb$_2$Te$_3$, and for $\theta_\acoustic$ in
the case of Al$_2$O$_3$. 
The Pearson correlation between the experimental and calculated
thermal conductivity values is $0.892$.
The corresponding Spearman correlation
is $0.600$.
The Spearman correlation between the experimental
values of the thermal conductivity and the values of $\theta_\acoustic$
calculated with \AGL\ is $0.943$.

\begin{table}[t!]
  \caption{\small Lattice thermal conductivity at 300K, Debye
    temperature and Gr{\"u}neisen parameter of rhombohedral
    semiconductors.
The experimental Debye temperatures are $\theta_\Debye$ for
Bi$_2$Te$_3$ and Sb$_2$Te$_3$, and  $\theta_\acoustic$ for Al$_2$O$_3$. 
   Units: $\theta$ in \K, $\kappa$ in \WmK.
}
 \label{tab:rhombo}
  \begin{tabular}{c c c c c c c c}
    \hline
    Comp. & $\theta^\EXP$ & $\theta_\acoustic^\AGL$ & $\theta_\Debye^\AGL$ & $\gamma^\EXP$ & $\gamma^\AGL$ & $\kappa^\EXP$  & $\kappa^\AGL$    \\
  %  &  \K &   \K &   &  &  \WmK  & \WmK   \\
 %   Compound & $\theta_\Debye^\EXP \K$  & $\theta_\Debye^\AGL$ \K & $\gamma^\EXP$ & $\gamma^\AGL$ & $\kappa^\EXP$   \WmK  & $\kappa^\AGL$   \WmK    \\
     \hline
    Bi$_2$Te$_3$ & 155 \cite{Snyder_jmatchem_2011} & 98 & 167 & 1.49 \cite{Snyder_jmatchem_2011} & 2.13 & 1.6 \cite{Snyder_jmatchem_2011} & 2.43 \\
    Sb$_2$Te$_3$ & 160 \cite{Snyder_jmatchem_2011} & 129 & 220 & 1.49 \cite{Snyder_jmatchem_2011} & 2.2 & 2.4 \cite{Snyder_jmatchem_2011} & 2.94 \\
    Al$_2$O$_3$ & 390 \cite{slack} & 376 & 810 & 1.32 \cite{slack} & 1.91 &  30 \cite{Slack_PR_1962} &  17.97  \\
    Cr$_2$O$_3$ &  & 262 & 565 &  & 2.26 & 16 \cite{Landolt-Bornstein, Bruce_PRB_1977} & 8.59 \\
    Fe$_2$O$_3$ &  & 182 & 388 &  & 5.32 & 11.3 \cite{Landolt-Bornstein, Horai_jgpr_1971} & 0.51 \\
    Bi$_2$Se$_3$ &  & 104 & 177 &  & 2.08 & 1.34 \cite{Landolt-Bornstein} & 2.88 \\
    \hline
  \end{tabular}  
\end{table}

The thermal conductivity of Fe$_2$O$_3$ is a clear outlier in this data set (see
fig.\ \ref{fig:rhombo}). Its Gr{\"u}neisen parameter, calculated with
Equation (\ref{miegruneisen}), is $5.32$. It is
abnormally high. Equation (\ref{gruneisen_theta}) gives a similar value of
$5.36$, whereas Equation (\ref{slatergruneisen}) gives a lower, but still
very high, value of $4.06$. 
Ignoring Fe$_2$O$_3$ in the comparison increases the
Pearson correlation of the calculated and experimental values of the
thermal conductivity to $0.992$, while the
Spearman correlation increases to $0.9$.

%\clearpage

\subsection{Body-centred tetragonal materials}

\begin{figure}[t!]
  \includegraphics[width=0.98\columnwidth]{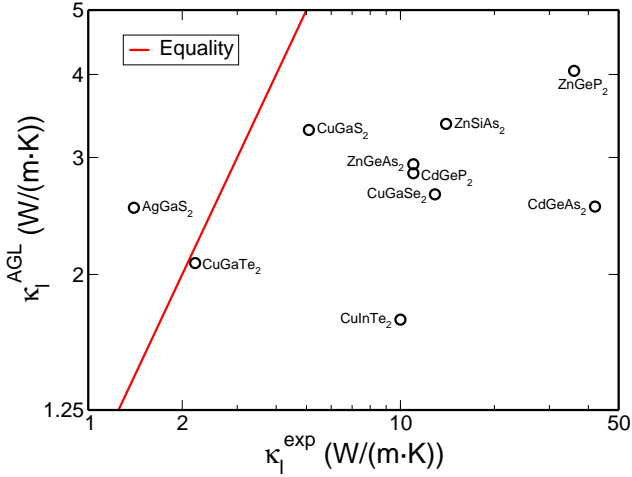}
  \vspace{-4mm}
  \caption{\small Lattice thermal conductivity of body-centred tetragonal
   semiconductors at 300K.}
  \label{fig:bct}
\end{figure}

Results for a set of body-centred tetragonal materials (spacegroup:
$I\overline{4}2d\ (\#122)$; Pearson symbol: $tI16$) are shown in Table \ref{tab:bct} and in
Figure  \ref{fig:bct}. For the materials ZnGeP$_2$ and AgGaS$_2$ there
are three and two experimental values listed for $\kappa^\EXP$. This is
due to the materials having different thermal conductivities in
different crystalline directions\cite{Beasley_AO_1994}. The following
results were obtained for the direction parallel to the optic axis,
36 W/(m$\cdot$K) and 1.4 W/(m$\cdot$K) for ZnGeP$_2$ and AgGaS$_2$,
respectively.
All of the experimental Debye temperatures listed in the table are 
the traditional Debye temperatures, $\theta_\Debye$.

\begin{table}[t!]
\caption{\small Lattice thermal conductivity at 300K, Debye
    temperature and Gr{\"u}neisen parameter of body-centred tetragonal
    semiconductors.
   Units: $\theta$ in \K, $\kappa$ in \WmK.
}
 \label{tab:bct}
  \begin{tabular}{c c c c c c c c}
   \hline
    Comp. & $\theta_\Debye^\EXP$ & $\theta_\acoustic^\AGL$ & $\theta_\Debye^\AGL$ & $\gamma^\EXP$ & $\gamma^\AGL$ & $\kappa^\EXP$  & $\kappa^\AGL$    \\
  %  &  \K &   \K &   &  &  \WmK  & \WmK   \\
 %   Compound & $\theta_\Debye^\EXP \K$  & $\theta_\Debye^\AGL$ \K & $\gamma^\EXP$ & $\gamma^\AGL$ & $\kappa^\EXP$   \WmK  & $\kappa^\AGL$   \WmK    \\
     \hline
    CuGaTe$_2$ & 226 \cite{Snyder_jmatchem_2011} & 141 & 281 & 1.46 \cite{Snyder_jmatchem_2011} & 2.32 & 2.2  \cite{Snyder_jmatchem_2011} & 2.08 \\
    ZnGeP$_2$ & 500 \cite{Landolt-Bornstein} & 176 & 351 & & 2.13 & 35 \cite{Landolt-Bornstein, Beasley_AO_1994} & 4.05 \\
     &  &  & & &  & 36 \cite{Landolt-Bornstein, Beasley_AO_1994} & \\
     &  &  & &  &  & 18 \cite{Landolt-Bornstein, Shay_1975, Masumoto_JPCS_1966} & \\
    ZnSiAs$_2$ & 347 \cite{Landolt-Bornstein, Bohnhammel_PSSa_1981} & 155 & 309 & & 2.15 & 14\cite{Landolt-Bornstein, Shay_1975, Masumoto_JPCS_1966} & 3.37 \\
    CuInTe$_2$ & 185 \cite{Landolt-Bornstein, Rincon_PSSa_1995} & 119 & 237 & 0.93 \cite{Rincon_PSSa_1995} & 2.33 & 10\cite{Landolt-Bornstein, Rincon_PSSa_1995} & 1.71 \\
     & 195 \cite{Landolt-Bornstein, Bohnhammel_PSSa_1982} & & &  &  &  & \\
    AgGaS$_2$ & 255 \cite{Landolt-Bornstein, Abrahams_JCP_1975} & 135& 269 & & 2.20 & 1.4\cite{Landolt-Bornstein, Beasley_AO_1994} & 2.52 \\
     & & & & & & 1.5\cite{Landolt-Bornstein, Beasley_AO_1994} & \\
    CdGeP$_2$ & 340 \cite{Landolt-Bornstein, Abrahams_JCP_1975} & 138 & 275 & & 2.20 & 11 \cite{Landolt-Bornstein, Shay_1975, Masumoto_JPCS_1966} & 2.84 \\
    CdGeAs$_2$ & & 140 & 280 & & 2.20 & 42 \cite{Landolt-Bornstein, Shay_1975} & 2.53 \\
    CuGaS$_2$ & 356 \cite{Landolt-Bornstein, Abrahams_JCP_1975} & 167 & 334 & & 2.24 & 5.09 \cite{Landolt-Bornstein} & 3.3 \\
    CuGaSe$_2$ & 262 \cite{Landolt-Bornstein, Bohnhammel_PSSa_1982} & 154 & 307 & & 2.27 & 12.9 \cite{Landolt-Bornstein, Rincon_PSSa_1995} & 2.64 \\
    ZnGeAs$_2$ & & 147 & 294 & & 2.16 & 11\cite{Landolt-Bornstein, Shay_1975} & 2.93 \\
    \hline
  \end{tabular}  
\end{table}

The Pearson correlation between the \AGL\
thermal conductivity values
and the experimental values is $0.383$. The corresponding Spearman correlation
is $0.498$. The Spearman correlation between the experimental
values of the thermal conductivity and the calculated values of $\theta_\acoustic$ is $0.401$.
The low correlations for this set of materials are
due to their anisotropic structure, where the materials display
different thermal conductivities along different lattice
directions. This demonstrates the limits of the isotropic
approximation made in the GIBBS method.

\subsection{Miscellaneous materials}

The results for materials with various other structures are shown in
Table \ref{tab:misc}. The materials are CoSb$_3$ and IrSb$_3$ (spacegroup:
$Im\overline{3}\ (\#204)$; Pearson symbol: $cI32$), ZnSb (spacegroup:
$Pbca\ (\#61)$; Pearson symbol: $oP16$), Sb$_2$O$_3$ (spacegroup:
$Pccn\ (\#56)$; Pearson symbol: $oP20$), InTe (spacegroup:
$Pm\overline{3}m\ (\#221)$; Pearson symbol: $cP2$), Bi$_2$O$_3$  (spacegroup:
$P121/c1\ (\#14)$; Pearson symbol: $mP20$), and SnO$_2$ (spacegroup:
$P42/mnm\ (\#136)$; Pearson symbol: $tP6$). The experimental Debye temperatures
listed in the table are 
the traditional Debye temperatures, $\theta_\Debye$, with the exception
of ZnSb for which it is $\theta_\acoustic$.

\begin{table}[t!]
  \caption{\small Lattice thermal conductivity at 300K, Debye
    temperature and Gr{\"u}neisen parameter of materials with various
    structures at 300K.
The experimental Debye temperatures are $\theta_\Debye$, 
except ZnSb for which it is $\theta_\acoustic$. 
  Units: $\theta$ in \K, $\kappa$ in \WmK.
}
  \label{tab:misc}
  \begin{tabular}{c c c c c c c c c}
   \hline
    Comp. & Pearson  & $\theta^\EXP$ & $\theta_\acoustic^\AGL$ & $\theta_\Debye^\AGL$ & $\gamma^\EXP$ & $\gamma^\AGL$ & $\kappa^\EXP$  & $\kappa^\AGL$    \\
 %  &  \K &   \K &   &  &  \WmK  & \WmK   \\
      \hline
    CoSb$_3$ & $cI32$ & 307 \cite{Snyder_jmatchem_2011} & 150 & 378 & 0.95 \cite{Snyder_jmatchem_2011} & 2.63 & 10 \cite{Snyder_jmatchem_2011} & 2.02 \\
    IrSb$_3$ & $cI32$ & 308 \cite{Snyder_jmatchem_2011} & 96 & 241 & 1.42 \cite{Snyder_jmatchem_2011} & 2.34 & 16 \cite{Snyder_jmatchem_2011} & 2.25 \\
    ZnSb & $oP16$ & 92 \cite{Madsen_PRB_2014} & 85 & 214 & 0.76 \cite{Madsen_PRB_2014} & 2.24 &  3.5 \cite{Madsen_PRB_2014, Bottger_JEM_2010} &  1.09  \\
    Sb$_2$O$_3$ & $oP20$ &  & 288 & 782 &  & 2.13 & 0.4 \cite{Landolt-Bornstein} & 6.07 \\
    InTe & $cP2$ & 186 \cite{Snyder_jmatchem_2011} & 152 & 191 & 1.0 \cite{Snyder_jmatchem_2011} & 2.28 & 1.7 \cite{Snyder_jmatchem_2011} & 3.12 \\
    Bi$_2$O$_3$ & $mP20$ &  & 85 & 232 &  & 2.1 & 0.8 \cite{Landolt-Bornstein} & 2.09 \\
    SnO$_2$ & $tP6$ &  & 515 & 935 &  & 2.48 & 98\cite{Turkes_jpcss_1980} & 15.0 \\
    &  &  & &  &  &   & 55 \cite{Turkes_jpcss_1980} & \\
    \hline
  \end{tabular}
\end{table}

For these materials, the Pearson  correlation between the calculated
and experimental values of the thermal conductivity is $0.914$. The corresponding
Spearman correlation is
$0.071$. The Spearman correlation between the experimental values
of the thermal conductivity and the calculated values of $\theta_\acoustic$ is
$0.143$. 

The low correlation values, particularly for the Spearman correlation,
for this set of materials demonstrates the
importance of the information about the material structure as an input
for the \AGL\ method. This is partly due to the fact that the Gr{\"u}neisen
parameter tends not to vary significantly between materials with a
particular structure, thus reducing its effect on the ordinal ranking of
the thermal conductivity of materials with the same structure.

\subsection{Half-Heusler materials}

\begin{figure}[t!]
%  \label{fig:Heusler_table_I}
  \includegraphics[width=0.98\columnwidth]{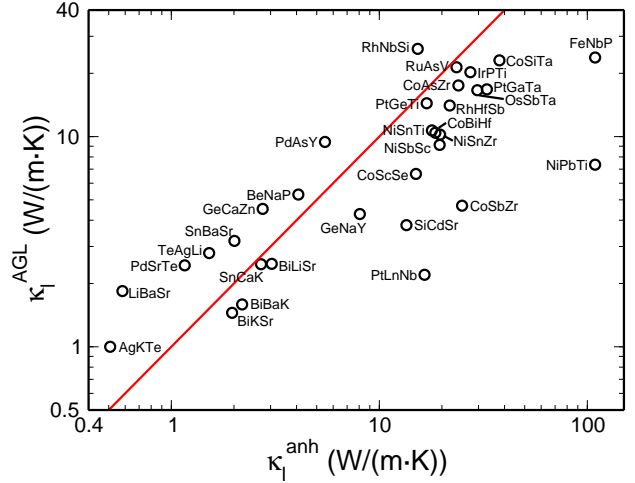}
  \vspace{-4mm}
  \caption{\small Thermal conductivities of half-Heusler semiconductors at
    300K compared to full anharmonic phonon {\it ab-initio}
    parameterization from Ref. \onlinecite{curtarolo:art84}.}
  \label{fig:Heusler_anh}
\end{figure}

Carrete et al. \cite{curtarolo:art84,curtarolo:art85}  studied the thermal
conductivity of 107 half-Heusler  (spacegroup: $F\overline{4}3m\ (\#216)$; Pearson symbol: $cF12$) compounds 
with {\it ab initio} and machine learning techniques.
In this section we compare their results with our \AGL\ calculations.
We first consider a subset of these half-Heusler materials, taken
from Table I of Ref. \onlinecite{curtarolo:art84}, for which the thermal
conductivity values were calculated using full anharmonic phonon
parameterization solutions of the BTE. 
The thermal conductivities at 300K for this set of materials as
calculated with 
Eq.\ (\ref{kappa_temperature}) are shown in Table \ref{tab:Heusler_anh} 
and in Figure   \ref{fig:Heusler_anh}. 
The Pearson correlation between the \AGL\ thermal conductivity
values and the full anharmonic phonon calculations is
$0.495$. The corresponding
Spearman correlation is $0.810$.
The Spearman correlation between the full anharmonic phonon
calculation values of the thermal conductivity and the values of $\theta_\acoustic$ as calculated with \AGL\ is $0.730$. 
A major contributor to the low Pearson correlation is the outlier
calculated value of the thermal conductivity of FeNbP and NiPbTi, 109.0
W/(m$\cdot$K) \cite{curtarolo:art84}.
If these materials are
removed from the dataset, the Pearson correlation increases to
$0.629$.

\begin{table}[t!]
  \caption{\small Thermal conductivities of half-Heusler semiconductors at 300K compared to full anharmonic phonon {\it ab-initio} parameterization from Ref. \onlinecite{curtarolo:art84}.
  Units: $\theta$ in \K, $\kappa$ in \WmK.
}
  \label{tab:Heusler_anh}
  \begin{tabular}{c c c c c c}
    \hline
    Comp. & $\theta_\acoustic^\AGL$ & $\theta_\Debye^\AGL$  & $\gamma^\AGL$ & $\kappa^\anh$ [\onlinecite{curtarolo:art84}] & $\kappa^\AGL$   \\
 %   &  \K &   &   \WmK  &  \WmK   \\
    \hline
    AgKTe & 105 & 152 & 2.26 &  0.508  & 1.0 \\
    BeNaP & 302 & 436 & 2.05 &  4.08  & 5.3 \\
    BiBaK & 95 & 137 & 1.94 &  2.19  & 1.59 \\
    BiKSr & 99 & 143 & 1.96 &  1.96  & 1.45 \\
    BiLiSr & 126 & 182 & 1.94 &  3.04  & 2.48 \\
    CoAsZr & 306 & 442 & 2.14 &  24.0  & 17.51 \\
    CoBiHf & 204 & 294 & 2.17 &  18.6  & 10.43 \\
    CoSbZr & 231 & 333 & 3.00 &  25.0  & 4.69 \\
    CoScSe & 230 & 331 & 2.09 &  15.0  & 6.64 \\
    CoSiTa & 296 & 427 & 1.92 &  37.8  & 23.06 \\
    FeNbP & 343 & 495 & 1.94 &  109.0  & 23.79 \\
    GeCaZn & 197 & 284 & 2.05 &  2.75  & 4.53 \\
    GeNaY & 189 & 273 & 2.04 &  8.06  & 4.28 \\
    LiBaSr & 88 & 127 & 1.33 &  0.582  & 1.84 \\
    IrPTi & 309 & 446 & 2.18 &  27.4  & 20.25 \\
    NiPbTi & 205 & 296 & 2.20 &  109.0  & 7.35 \\
    NiSbSc & 232 & 334 & 2.03 &  19.5  & 9.13 \\
    NiSnTi & 249 & 359 & 2.06 &  17.9  & 10.7 \\
    NiSnZr & 229 & 330 & 2.06 &  19.6  & 10.22 \\
    OsSbTa & 227 & 328 & 2.14 &  29.6  & 16.62 \\
    PdAsY & 230 & 332 & 2.17 &  5.48  & 9.43 \\
    PdSrTe & 130 & 188 & 2.13 &  1.16  & 2.44 \\
    PtGaTa & 242 & 349 & 2.19 &  32.9  & 16.78 \\
    PtGeTi & 263 & 379 & 2.23 &  16.9  & 14.41 \\
    PtLaNb & 140 & 202 & 2.69 &  16.5  & 2.2 \\
    RhHfSb & 232 & 335 & 2.18 &  21.8  & 14.06 \\
    RhNbSi & 345 & 497 & 2.09 &  15.3  & 26.15 \\
    RuAsV & 334 & 482 & 2.19 &  23.5  & 21.37 \\
    SbCaK & 141 & 203 & 1.92 &  2.70  & 2.47 \\
    SiCdSr & 168 & 242 & 2.05 &  13.5  & 3.79 \\
    SnBaSr & 114 & 165 & 1.71 &  2.01  & 3.19 \\
    TeAgLi & 166 & 239 & 2.32 &  1.52  & 2.79 \\
    \hline
  \end{tabular}
\end{table}

\begin{figure}[h!]
  \includegraphics[width=0.98\columnwidth]{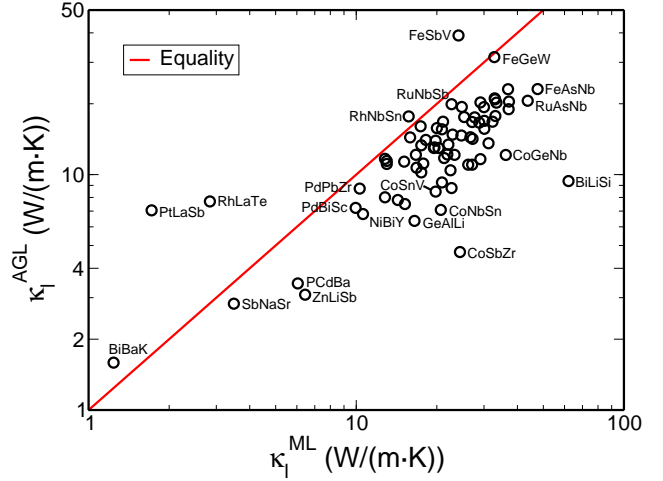}
  \vspace{-4mm}
  \caption{\small Thermal conductivities of half-Heusler semiconductors at
    300K compared to machine learning algorithm predictions from
    Ref. \onlinecite{curtarolo:art84}.}
  \label{fig:Heusler_ML}
\end{figure}

The second subset of half-Heusler materials studied is taken
from Table III of Ref. \onlinecite{curtarolo:art84}, 
where the thermal conductivity was estimated using a machine learning algorithm.
Comparison of these values with the thermal conductivity at 300K 
calculated with Eq.\
(\ref{kappa_temperature}) is shown in
Table \ref{tab:Heusler_ML} and Figure   \ref{fig:Heusler_ML}.
The Pearson correlation between the \AGL\ thermal conductivities and
those produced by the machine learning algorithm is $0.578$. The
corresponding Spearman correlation
is $0.706$. The Spearman correlation between the machine learning
thermal conductivities and the \AGL\ values of $\theta_\acoustic$ is $0.679$.

%\begin{widetext}

% \begin{longtable}[t!]
\begin{table*}[t!]
  \caption{\small Thermal conductivities of half-Heusler
    semiconductors at 300K compared to machine learning algorithm
    predictions from Ref. \onlinecite{curtarolo:art84}. Units: $\theta$ in \K, $\kappa$ in \WmK.}
  \label{tab:Heusler_ML}
%\begin{table}[h!]
  \begin{tabular}{c c c c c c | c c c c c c}
    \hline
    Comp. & $\theta_\acoustic^\AGL$ & $\theta_\Debye^\AGL$  & $\gamma^\AGL$ & $\kappa^\ML$ [\onlinecite{curtarolo:art84}]  & $\kappa^\AGL$ & Comp. & $\theta_\acoustic^\AGL$ & $\theta_\Debye^\AGL$   & $\gamma^\AGL$ & $\kappa^\ML$ [\onlinecite{curtarolo:art84}]  & $\kappa^\AGL$    \\
%    &   \K &  &   \WmK  &   \WmK   &    &   \K &  &   \WmK  &   \WmK   \\
    \hline
    AuAlHf & 217 & 313 & 2.12 &  16.7  & 12.14 & NiBiSc & 207 & 299 & 2.17 &  14.3  & 7.8 \\
    BLiSi & 433 & 624 & 2.07 &  62.1  & 9.39 & NiBiY & 187 & 269 & 2.16 &  10.6  & 6.8 \\
    BiBaK & 95 & 137 & 1.94 &  1.24  & 1.59 & NiGaNb & 289 & 417 & 2.11 &  22.9  & 14.79 \\
    CoAsHf & 266 & 383 & 2.13 &  20.0  & 15.76 & NiGeHf & 238 & 343 & 2.02 &  19.6  & 13.05 \\
    CoAsTi & 345 & 497 & 2.15 & 37.1  & 18.96 & NiGeTi & 330 & 476 & 2.10 &  25.3  & 17.56 \\
    CoAsZr & 306 & 442 & 2.14 &  27.7  & 17.51 & NiGeZr & 295 & 426 & 2.07 &  21.1  & 16.78 \\
    CoBiHf & 204 & 294 & 2.17 &  22.5  & 10.43 & NiHfSn & 230 & 332 & 2.08 &  19.5  & 12.97 \\
    CoBiTi & 236 & 341 & 2.19 &  27.1  & 11.02 & NiPbZr & 195 & 281 & 2.19 &  15.2  & 7.5 \\
    CoBiZr & 223 & 322 & 2.17 &  17.8  & 11.14 & NiSnTi & 249 & 359 & 2.06 &  16.8  & 10.7 \\
    CoGeNb & 295 & 425 & 2.39 &  36.2  & 12.12 & NiSnZr & 229 & 330 & 2.06 &  17.5  & 10.22 \\
    CoGeTa & 266 & 383 & 2.24 &  27.2  & 14.19 & OsNbSb & 254 & 367 & 2.12 &  24.8  & 19.38 \\
    CoGeV & 334 & 482 & 2.01 &  29.1  & 20.26 & OsSbTa & 227 & 328 & 2.14 &  28.8  & 16.62 \\
    CoHfSb & 190 & 274 & 1.69 &  21.9  & 12.18 & PCdBa & 198 & 285 & 2.24 &  6.05  & 3.45 \\
    CoNbSi & 323 & 466 & 2.18 &  30.1  & 15.65 & PdBiSc & 194 & 280 & 2.23 &  9.95  & 7.22 \\
    CoNbSn & 238 & 343 & 2.56 &  20.7  & 7.08 & PdGeZr & 267 & 385 & 2.18 &  18.2  & 14.04 \\
    CoSbTi & 263 & 379 & 2.10 &  23.3  & 12.13 & PdHfSn & 218 & 314 & 2.21 &  15.1  & 11.35 \\
    CoSbZr & 231 & 333 & 3.0 &  24.4  & 4.69 & PdPbZr & 203 & 293 & 2.29 &  10.3  & 8.72 \\
    CoSiTa & 296 & 427 & 1.92 &  36.9  & 23.06 & PtGaTa & 242 & 349 & 2.19 &  32.3  & 16.78 \\
    CoSnTa & 217 & 313 & 2.32 &  22.7  & 8.77 & PtGeTi & 263 & 379 & 2.23 &  26.7  & 14.41 \\
    CoSnV & 266 & 383 & 2.49 &  19.8  & 8.47 & PtGeZr & 245 & 354 & 2.19 &  15.9  & 14.39 \\
    FeAsNb & 339 & 489 & 2.13 &  47.6  & 23.09 & PtLaSb & 168 & 243 & 2.11 &  1.72  & 7.05 \\
    FeAsTa & 295 & 425  & 2.13 & 32.9  & 21.08 & RhAsTi & 311 & 449 & 2.18 &  33.1  & 17.74 \\
    FeGeW & 245 & 354 & 1.40 &  32.8  & 31.46 & RhAsZr & 284 & 409 & 2.17 &  27.1  & 16.73 \\
    FeNbSb & 216 & 311 & 1.79 &  29.1  & 11.63 & RhBiHf & 182 & 263 & 2.25 &  12.8  & 8.01 \\
    FeSbTa & 196 & 282 & 1.65 &  31.2  & 13.59 & RhBiTi & 228 & 329 & 2.25 &  13.0  & 11.1 \\
    FeSbV & 305 & 440 & 1.50 &  24.1  & 39.0 & RhBiZr & 218 & 314 & 2.22 &  13.0  & 11.43 \\
    FeTeTi & 266 & 384 & 2.24 &  26.2  & 11.02 & RhLaTe & 195 & 281 & 2.24 &  2.84  & 7.69 \\
    GeAlLi & 270 & 390 & 2.06 &  16.5  & 6.36 & RhNbSn & 275 & 396 & 2.19 &  15.7  & 17.67 \\
    IrAsTi & 277 & 399 & 2.22 &  30.1  & 16.92 & RhSnTa & 227 & 327 & 2.18 &  20.3  & 12.98 \\
    IrAsZr & 255 & 368 & 2.19 &  17.4  & 16.04 & RuAsNb & 306 & 442 & 2.17 &  43.7  & 20.59 \\
    IrBiZr & 206 & 297 & 2.24 &  12.8  & 11.67 & RuAsTa & 279 & 402 & 2.19 &  33.4  & 20.21 \\
    IrGeNb & 279 & 402 & 2.17 &  33.0  & 20.88 & RuNbSb & 284 & 409 & 2.17 &  22.7  & 19.91 \\
    IrGeTa & 256 & 369 & 2.15 &  37.2  & 20.43 & RuSbTa & 239 & 344 & 2.15 &  20.9  & 15.58 \\
    IrGeV & 288 & 416 & 2.19 &  30.0  & 19.34 & RuTeZr & 241 & 348 & 2.26 &  21.3  & 11.76 \\
    IrHfSb & 221 & 319 & 2.20 &  24.7  & 14.66 & SbNaSr & 139 & 200 & 1.90 &  3.49  & 2.83 \\
    IrNbSn & 232 & 334 & 2.18 &  19.8  & 13.93 & SiAlLi & 363 & 523 & 2.02 &  20.9  & 9.26 \\
    IrSnTa & 218 & 314 & 2.23 &  22.1  & 13.42 & ZnLiSb & 176 & 254 & 2.12 &  6.44  & 3.09 \\
    NiAsSc & 300 & 432 & 2.11 &  17.5  & 13.32 & & & & & & \\
    \hline
  \end{tabular}
  % \end{longtable}
\end{table*}
%\end{widetext}

\begin{table}[ht]
%\begin{table*}[ht!]
  \caption{\small Thermal conductivities of half-Heusler
    semiconductors at 300K compared to experimental values.  Units: $\theta$ in \K, $\kappa$ in \WmK.}
  \label{tab:Heusler_exp}
%\begin{table}[h!]
  \begin{tabular}{c c c c c c c}
    \hline
    Comp. & $\theta_\acoustic^\AGL$ & $\theta_\Debye^\AGL$  & $\gamma^\AGL$ & $\kappa^\ML$  [\onlinecite{curtarolo:art84}]  & $\kappa^\AGL$ & $\kappa^\EXP$   \\
%    Comp. & $\theta$   & $\gamma$ & $\kappa$   & $\kappa$ & exp & Comp. & $\theta$   & $\gamma$ & $\kappa$   & $\kappa$ & exp    \\
%     &   \K &  &   \WmK  &   \WmK   &    &   \K &  &   \WmK  &   \WmK   \\
    \hline
    CoHfSb & 190 & 274 & 1.69 &  21.9  & 12.18 & 17\cite{Sekimoto2005} \\
    CoSbTi & 263 & 379 & 2.10 &  23.3  & 12.13  & 12\cite{Kawaharada2004}  \\
     & & & & & & 25\cite{Xia2000} \\
    CoSbZr & 231 & 333 & 3.0 &  24.4  & 4.69  & 15\cite{Xia2000}    \\
    FeSbV & 305 &  440 & 1.50 &  24.1  & 39.0  & 13\cite{Young2000} \\
    NiHfSn & 230 & 332 & 2.08 &  19.5  & 12.97  & 6.7\cite{Hohl1999} \\
    NiSnTi & 249 & 359 & 2.06 &  16.8  & 10.7  & 9.3\cite{Hohl1999} \\
    NiSnZr & 229 & 330 & 2.06 &  17.5  & 10.22 & 8.8\cite{Hohl1999} \\
    & & & & & & 17.2\cite{uher1999} \\
    \hline
  \end{tabular}
  % \end{longtable}
\end{table}

\begin{figure}[h!]
  \includegraphics[width=0.98\columnwidth]{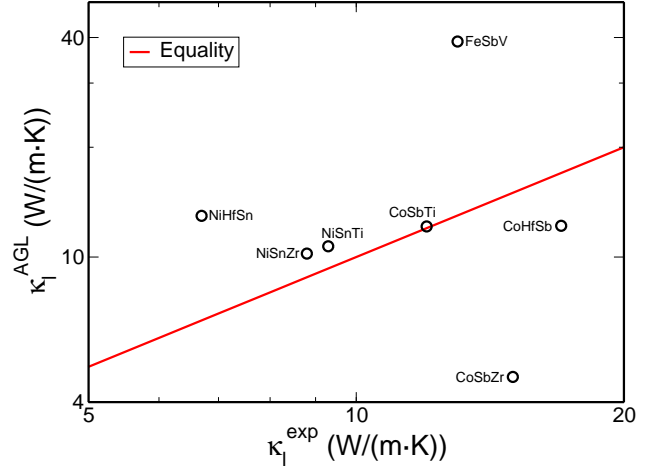}
  \vspace{-4mm}
  \caption{\small Thermal conductivities of half-Heusler semiconductors at
    300K compared to experimental measurements.}
  \label{fig:Heusler_exp}
\end{figure}

Experimental results for the thermal conductivity of 7 of these
half-Heusler materials were available in the
literature, and these values are shown in Table \ref{tab:Heusler_exp}
and Figure \ref{fig:Heusler_exp}.  The Pearson correlation between the \AGL\ thermal conductivities and
the experimental values is $0.064$, while the
corresponding Spearman correlation
is $-0.036$. The Spearman correlation between the experimental
thermal conductivities and the \AGL\ values of $\theta_\acoustic$ is
$0.0$. However, this is a small sample set, and these low correlation values appear to be primarily due to the
outlier material CoSbZr, for which \AGL\ predicts a relatively high
value of $3.0$ for the Gr{\"u}neisen parameter. Ignoring this material
in the comparison increases the Pearson correlation between the
thermal conductivities to $0.262$ and the Spearman correlation to $0.314$.

\subsection{\AGL\ predictions for zincblende materials}

\begin{figure}[h!]
  \includegraphics[width=0.98\columnwidth]{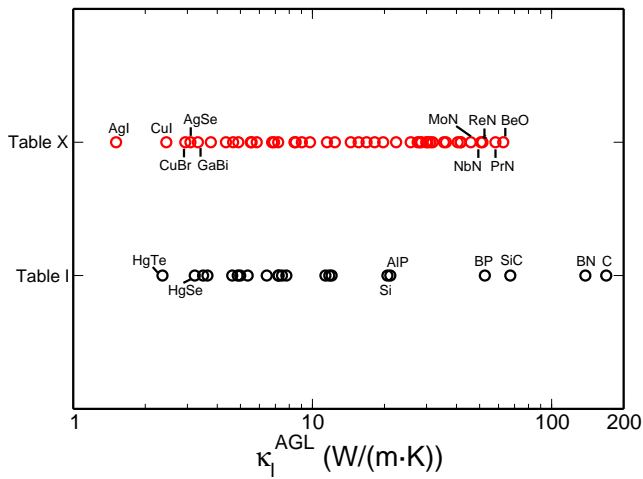}
  \vspace{-4mm}
  \caption{\small Predicted thermal conductivities of zincblende
    structure materials at 300K. The \AGL\ values for the materials with experimental
    data listed in Table \ref{tab:zincblende}
    are shown in black (see also figure \ref{fig:zincblende}).
    \AGL\ predictions for materials with no experimental data are in red.}
  \label{fig:zincblende_prediction}
\end{figure}

In order to demonstrate the potential utility of the \AGL\ method for
high-throughput screening of the thermal properties of materials
we have calculated the Debye temperature, Gr{\"u}neisen parameter and
thermal conductivity for 45 zincblende structure (spacegroup: $F\overline{4}3m\ (\#216)$; Pearson symbol: $cF8$) materials which were not included
in Table \ref{tab:zincblende}, and for which experimental values of
the thermal conductivity do not seem to be available in the
literature. The results for these materials are shown in Table
\ref{tab:zincblende_prediction} and in figure
\ref{fig:zincblende_prediction}. 

From these results, it is noticeable that BeO is predicted to have the
highest thermal conductivity, with a value similar to that of SiC. This
high thermal conductivity is in agreement with recent first principles
calculations \cite{Li_JAP_BeO_2013}. Another set of
materials predicted to have high thermal conductivity
includes the nitrides PrN, ReN, NbN and MoN. Although BAs was previously predicted to have an extremely
high thermal conductivity \cite{Lindsay_PRL_2013}, the \AGL\ value is
only slightly higher than that of Si or AlP, and less than that of BP,
BN or SiC.
The materials with the lowest thermal conductivity in this set are
AgI and CuI. AgI, in particular, is predicted by \AGL\ to have a thermal
conductivity lower than that of any of the materials in Table \ref{tab:zincblende}. 

\begin{table}[ht!]
 \caption{\small Lattice thermal conductivity at 300K, Debye
   temperature and Gr{\"u}neisen parameter of Zincblende structure
   materials for which the experimental thermal conductivity is not
   available in the literature. 
  Units: $\theta$ in \K, $\kappa$ in \WmK.
 }
 \label{tab:zincblende_prediction}
% {\footnotesize
 \begin{tabular}{c c c c c | c c c c c}
  \hline
  Comp. & $\theta_\acoustic^\AGL$ & $\theta_\Debye^\AGL$ & $\gamma^\AGL$ & $\kappa^\AGL$ & Comp. & $\theta_\acoustic^\AGL$ & $\theta_\Debye^\AGL$ & $\gamma^\AGL$ & $\kappa^\AGL$    \\
  %  &  \K &   \K &   &  &  \WmK  & \WmK   \\
%    Compound & $\theta_\Debye^\EXP \K$  & $\theta_\Debye^\AGL$ \K & $\gamma^\EXP$ & $\gamma^\AGL$ & $\kappa^\EXP$   \WmK  & $\kappa^\AGL$   \WmK    \\
     \hline
    AgI & 123 & 155 & 2.45 & 1.51 & MgSe & 250 & 315 & 1.91 & 8.51 \\
    AgO & 247 & 311 & 2.06 & 7.19 & MgTe & 187 & 236 & 1.94 & 5.57 \\
    AgSe & 179 & 225 & 2.52 & 3.09 & MoN & 447 & 563 & 1.82 & 45.92  \\
    AuN & 259 & 326 & 2.49  & 9.03 & NbN & 493 & 621 & 1.97 & 50.62 \\
    BAs & 420 & 529 & 1.95 & 25.75 & NiN & 416 & 524 & 1.58 & 30.30 \\
    BeO & 845 & 1065 & 1.74 & 62.77 & PdN & 387 & 487 & 2.37 & 18.2 \\
    BeS & 506 & 637 & 1.76 &  27.54 & PrN & 309 & 389 & 1.27 & 58.21 \\
    BeSe & 324 & 408 & 1.80 & 15.6 & ReN & 385 & 485 & 1.83 & 51.39 \\
    BeTe & 237 & 299 & 1.85 & 9.77 & RhN & 450 & 567 & 2.27 & 29.91 \\
    CaSe & 208 & 262 & 1.84 & 6.78 & RuN & 487 & 614 & 2.18 & 40.49 \\
    CdS & 228 & 287 & 2.15 & 6.91 & SbSn & 143 & 180 & 1.70 & 5.52 \\
    CoO  & 427 & 538 & 2.41 & 14.46 & ScSi & 298 & 375 & 1.78 & 12.41 \\
    CuBr & 190 & 239 & 2.44 & 2.94 & ScSn & 177 & 223 & 1.85 & 5.85 \\
    CuCl & 234 & 295 & 2.40 & 3.76 & SiP & 357 & 450 & 2.97 & 4.90 \\
    CuF & 272 & 343 & 2.34 & 4.66 & TaN & 379 & 477 & 1.97 & 41.64 \\
    CuI & 161 & 203 & 2.48 & 2.45 & TcB & 371 & 468 & 1.59 & 35.46 \\
    GaBi & 140 & 177 & 2.18 & 3.32 & TcN & 469 & 591 & 2.43 & 28.08 \\
    GdO & 275 & 346 & 1.92 & 16.84 & TiB & 448 & 565 & 1.69 & 31.06 \\
    GeP & 239 & 301 & 1.54 & 11.5 & WN & 344 & 433 & 2.44 & 19.76 \\
    GeSc & 231 & 291 & 1.88 & 8.41 & YN & 373 & 470 & 1.80 & 28.51 \\
    HfN & 348 & 439 & 1.89 & 36.14 & ZnO & 417 & 525 & 1.95 & 22.38 \\
    HgS & 176 & 222 & 2.34 & 4.35 & ZrN & 450 & 567 & 1.92 & 41.65 \\
    IrN & 371 & 467 & 2.19 & 31.79 & \\
    \hline
  \end{tabular}
%}
\end{table}

\begin{table}[t!]
\caption{\small Summary of correlations between experimental and  \AGL\ lattice thermal conductivity values. The total value is
  for the set containing all of the non-half Heusler materials.}
 \label{tab:kappa_correlation}
  \begin{tabular}{c c c c}
    \hline
    Comp. set & Pearson & Spearman & Spearman   \\
    &  $\kappa^\EXP\!\!\leftrightarrow\!\kappa^\AGL$ & $\kappa^\EXP\!\!\leftrightarrow\!\kappa^\AGL$ & $\kappa^\EXP\!\!\leftrightarrow\!\theta_\acoustic^\AGL$   \\
    \hline
    Zincblende & 0.878 & 0.905 & 0.925 \\
    Rocksalt & 0.910 & 0.445 & 0.645 \\
    Wurzite & 0.943 & 0.976 & 0.905 \\
    Rhombohedral & 0.892 & 0.600 & 0.943 \\
    Tetragonal & 0.383 & 0.498 & 0.401 \\
    Misc. &  0.914 & 0.071 & 0.143 \\
    Total & 0.879 & 0.730 & 0.736 \\
    \hline
  \end{tabular}  
\end{table}

\begin{table}[t!]
\caption{\small Summary of correlations between {\it ab initio} and  \AGL\ lattice thermal conductivity values for the half Heusler materials.}
 \label{tab:Heusler_kappa_correlation}
  \begin{tabular}{c c c c}
    \hline
    Comp. set & Pearson & Spearman & Spearman   \\
    &  $\kappa^\anh\!\!\leftrightarrow\!\kappa^\AGL$ & $\kappa^\anh\!\!\leftrightarrow\!\kappa^\AGL$ & $\kappa^\anh\!\!\leftrightarrow\!\theta_\acoustic^\AGL$   \\
    \hline
    Full anharmonic & 0.495 & 0.810 & 0.730 \\
    Machine learning & 0.578 & 0.706 & 0.679 \\
    \hline
  \end{tabular}  
\end{table}

\newpage
\newpage

\section{Conclusions}

We implemented the ``GIBBS'' quasi-harmonic Debye model in the
\AGL\  software package within
the  {\small AFLOW} and Materials Project high-throughput
computational materials science frameworks. We used it 
to  automatically calculate the thermal conductivity, Debye
temperature and Gr{\"u}neisen coefficient of materials with
various structures and compared them with experimental results.

A major aim of high-throughput calculations is
to identify useful markers (descriptors) for screening large datasets of structures
for desirable properties \cite{curtarolo:art81}.
In this study we examined whether the {\it inexpensive-to-calculate}
Debye model thermal properties may be useful as such
markers for high thermal conductivity materials, despite the well
known deficiencies of this model in their quantitative evaluation.
We therefore concentrated on correlations between the calculated
quantities and the corresponding experimental data.
  
The correlations between the experimental values of the thermal
conductivity and those calculated with  \AGL\ are summarized in
Table \ref{tab:kappa_correlation}. For the entire set of materials
examined we find a high 
Pearson correlation of $0.879$ between $\kappa^\EXP$  and
$\kappa^\AGL$. It is particularly high, above $0.9$, for materials
with high symmetry (cubic or rhombohedral) structures, but
significantly lower for anisotropic materials.
We also compared these results with similar calculations of the thermal
conductivity, using the experimental values of the Debye
temperature and Gr{\"u}neisen coefficient. The two methods gave
similar Pearson correlations for the thermal conductivities,
demonstrating that the  \AGL\ approach  can rectify the lack of
this experimental data in screening
large data sets of materials.

The Spearman correlation between $\kappa^\EXP$ and
$\theta^\AGL$ for the entire set of materials is almost as high as the
Pearson correlation between the calculated and experimental
conductivities. It is, however, less consistent for the high symmetry
structures, with  a relatively low value of $0.645$ for the Rocksalt
structures.
The Spearman correlation between $\kappa^\EXP$ and
$\kappa^\AGL$ is found to be inferior to both previous measures as
a descriptor of high conductivity materials.
The correlations for the half-Heusler materials are summarized in
Table \ref{tab:Heusler_kappa_correlation}.

Overall, despite the quantitative limitations of the method, the \AGL\ approach can be useful for quickly
screening large data sets of materials for favorable thermal properties.

\section{Acknowledgments}
We thank Drs. Jesus Carrete, Natalio Mingo, Gus Hart, Anubhav Jain, Shyue Ping
Ong, Kristin Persson, and Gerbrand Ceder for various technical discussions.
We acknowledge support by the DOE (DE-AC02- 05CH11231), specifically the Basic Energy Sciences program under Grant \# EDCBEE. 
The consortium {\small AFLOWLIB}.org acknowledges Duke University -- Center for Materials Genomics --- and the CRAY corporation for computational support.

%\bibliographystyle{PhysRevwithTitles_noDOI_v1b} %EOPAPER
%\bibliographystyle{unsrt}
%\normalbaselines %Fixes spacing of bibliography
%\bibliography{xstefano4,xcormac} %your bibliography

\end{document}